\documentclass[a4paper,12pt]{article}

\usepackage[utf8x]{inputenc}
\usepackage{amsmath,amssymb}
\usepackage{tikz}
\usepackage{graphicx}
\usepackage{cite}
\numberwithin{equation}{section}

\usepackage[singlelinecheck=false,justification=justified]{caption}

\usepackage{esint}

\usepackage{lipsum}
\usepackage{lmodern}
\usepackage{tcolorbox}

\usepackage{empheq}
\usepackage{mathtools}

\usepackage{cancel} 

\newcommand{\braket}[1]{\langle\hskip -1mm \langle #1 \rangle\hskip -1mm\rangle}
\newcommand{\beq}{\begin{equation}\begin{aligned}{}}
\newcommand{\eeq}{\end{aligned}\end{equation}}

\graphicspath{ {images/} }
 \usepackage[toc,page]{appendix}

\usepackage{wrapfig}
\usepackage{float}
\usepackage{xcolor}
\usepackage{amsmath}
\usepackage{amsfonts}

 \usepackage{hyperref}
\hypersetup{
    colorlinks=true,
    citecolor=blue,
    linkcolor=blue,
    filecolor=magenta,
    urlcolor=blue}

 \newcommand{\bea}{\begin{eqnarray}}
\newcommand{\eea}{\end{eqnarray}}

\newcommand{\abs}[1]{\lvert #1\rvert}

\renewcommand{\Re}{\operatorname{Re}}

\newsavebox{\uuunit}
\sbox{\uuunit}
    {\setlength{\unitlength}{0.825em}
     \begin{picture}(0.6,0.7)
        \thinlines
        \put(0,0){\line(1,0){0.5}}
        \put(0.15,0){\line(0,1){0.7}}
        \put(0.35,0){\line(0,1){0.8}}
       \multiput(0.3,0.8)(-0.04,-0.02){10}{\rule{0.5pt}{0.5pt}}
     \end {picture}}

\renewcommand{\cal}[1]{\mathcal{#1}}

\begin{document}
 

\title{\bf  Invariant tensions  from holography
  \vskip 5mm
\,  }

\author{Constantin Bachas  and Zhongwu Chen }


\maketitle
\begin{center}
\textit{Laboratoire de Physique  de l'\'Ecole Normale Sup\'eri{e}ure,\\
CNRS, PSL  Research University  and Sorbonne Universit\'es \\
24 rue Lhomond, 75005 Paris, France}
\end{center}

\vskip 1.5cm

\begin{abstract}

We  revisit the problem of defining an invariant notion of  tension  in gravity. 
For spacetimes whose asymptotics are those of a Defect CFT we propose 
  two independent definitions :
Gravitational tension given by the one-point function of the dilatation current, and
 inertial tension, or stiffness, given by  the norm of the displacement  operator.
We show that  both reduce to the tension of the Nambu-Goto action in the limit of classical  thin-brane probes. 
Subtle normalisations of  the relevant
 Witten diagrams are fixed by 
 the Weyl and diffeomorphism Ward identities of the  boundary DCFT. 
The gravitational tension is not defined for domain walls, whereas  stiffness is not defined for point particles. 
When they both exist
these  two   tensions  are in general different, 
but the examples of line and surface BPS defects in $d=4$ show that superconformal  invariance can 
identify them. 

\end{abstract}

\newpage

\section{\large Introduction}

   A key entry in the AdS/CFT dictionary \cite{Maldacena:1997re,Gubser:1998bc,Witten:1998qj}
     is the relation between the mass, $m_0$, of a point particle in the bulk  and the 
     scaling dimension,  $\Delta$,  of the  dual  CFT operator. For a free scalar particle in AdS$_{d+1}$  
     \bea\label{Delta}
      \Delta\, =\, {d\over 2} \, + \, \sqrt{(m_0{\ell})^2 + {d^2\over 4}} \, = \, m_0{\ell} + {d\over 2} +  {d^2 \over 8m_0{\ell}} + \cdots\  
     \eea
where ${\ell}$ is the AdS radius which we set henceforth   equal to one.  
In the world-line formalism  one finds $\langle {\cal O}(x) {\cal O}(y)\rangle  \sim e^{-m_0\, {\mathfrak D}}$,  where ${\cal O}$ is the  dual CFT operator
inserted at the points $x$ and $y$ on the Poincar\'e boundary,    
 ${\mathfrak D} = 2 \log(\vert x-y\vert/\epsilon)$ is the geodesic distance between these  points and $\epsilon$ is the boundary cutoff. 
This explains the leading term in the expansion \eqref{Delta}. 
The next term  comes from Gaussian quantum fluctuations,\footnote{Which in general  depend   on the particle's spin.}
 and subleading  ones from the non-linearities of the point-particle action 
 which are  conveniently resummed by the Klein-Gordon equation.
 
    Such corrections  are negligible if the Compton wavelength of the particle is much smaller than the  AdS radius,\,  $ m_0^{-1}  \ll 1$. 
In addition,  for the  point-particle description to stay valid the AdS radius must  be much larger  than the Schwarzschild radius, $r_S \sim G_N m_0\ll 1$.  
 In the language of  ref.\cite{Abajian:2023jye} one may refer to  the  range $1\ll m_0 \ll 1/G_N$ as that of {\it heavy}  but not {\it huge} 
 operators. The   latter correspond to black-hole micro-states for which the 
 world-line approximation breaks down and the geometry thickens,  as illustrated below. 
 
\includegraphics[width=0.83\textwidth]{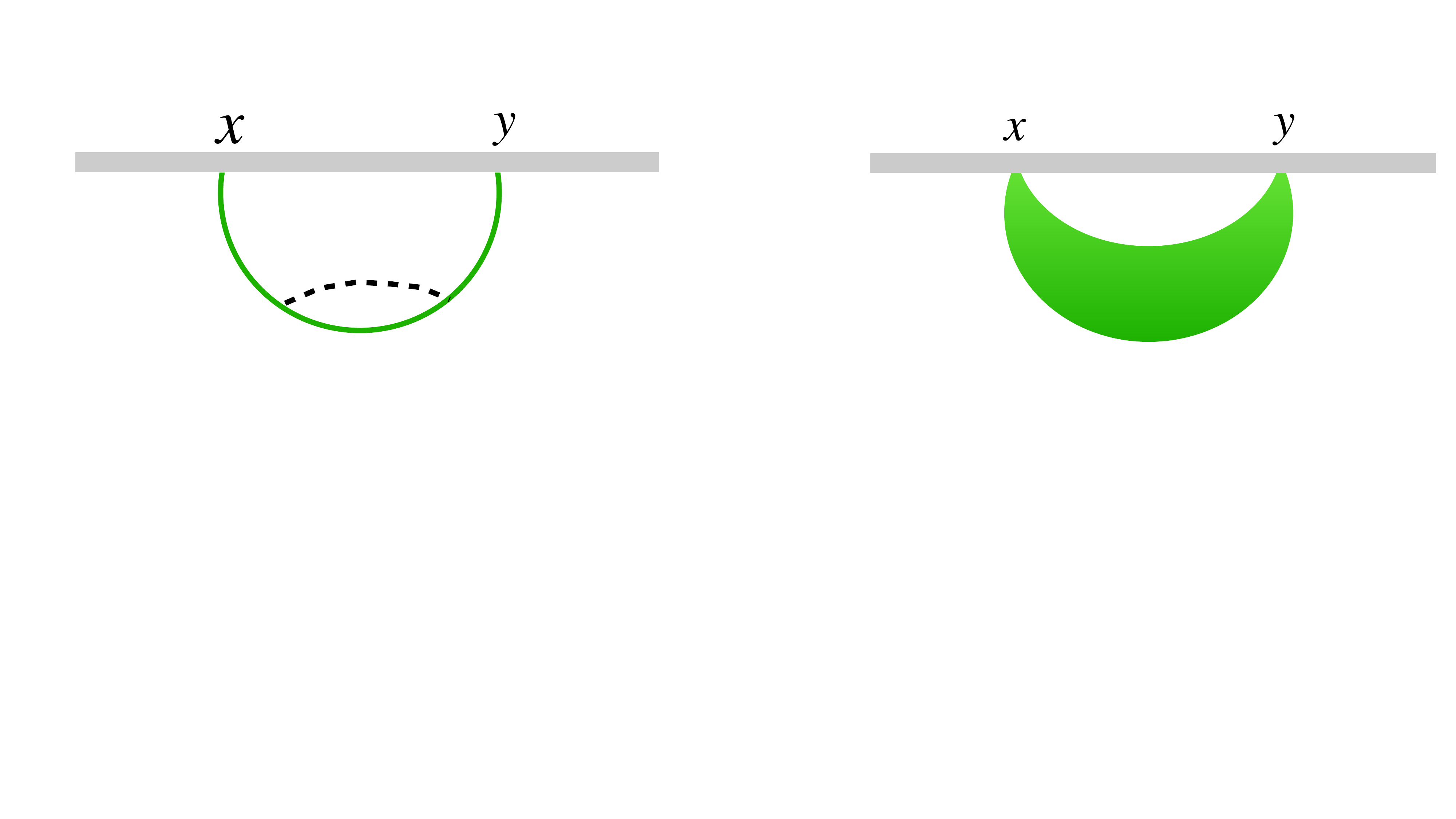}
\vskip  -2.9cm
  \centerline{Figure 1: \footnotesize  A `banana-shaped'  geometry (right)  replaces the geodesic world-line (left)
 }
 \leftline{ \footnotesize  \  when $G_n m \gtrapprox 1$\,. The AdS boundary is in grey. The broken line on the left is   a virtual  \\ 
}
 \leftline{ \footnotesize\   graviton that screens the bare mass parameter  of the point-particle action.
}
 
\vskip 3mm

 \noindent 
  Whether or not the dual `banana-shaped' geometry  is   a  smooth  horizonless  fuzzball  
\cite{Mathur:2005zp}, a bound state  or  a black hole,  mass cannot be defined locally anymore. 
In the world-line formalism virtual gravitons screen the bare mass parameter  $m_0$. 
But an invariant energy does exist  in global AdS \cite{Abbott:1981ff,Hawking:1995fd},  and it coincides (up to a Casimir subtraction) with the dilatation charge
$\Delta$.  The existence of this invariant charge  makes it  possible to count  the microscopic  bulk states in the dual CFT.

  
All this is well known. The question that we would like  to address here  is 
 whether  the holographic dictionary contains  a similar entry  for the tension   of  $p$-dimensional branes.
A brane in AdS can be compact, in which case its only gravitational charge is  energy, but it may also have infinite extent and 
  intersect the  boundary along  a $p$-dimensional defect
  \cite{Maldacena:1998im,Karch:2000gx,DeWolfe:2001pq,Bachas:2001vj}.  It should in this case be possible to give 
 an invariant  definition of tension  on the AdS boundary. When  a dual Defect Conformal Field Theory (DCFT) exists, the definition should only depend
 on its data.\footnote{But  
     the  existence of such a dual is not necessary -- 
  DCFT language is just  a proxy for  the asymptotic AdS data.} 
 As for point particles, it must also reduce to the  bare  Nambu-Goto tension, $T_p^{(0)}$,  in the 
  limit of a thin,  heavy but not huge  brane (i.e. when $1\ll T_p^ {(0)} \ll G_N T_p^{(0)}$).  
We will argue that  two natural candidates  fit the bill:


 \smallskip
  ({\footnotesize I}) The integrated one-point function of the dilatation current, 
\bea\label{def1}
\boxed{\, T_p^{\,(\footnotesize\rm I)} \, := \,  \bigl({\,\scalebox{0.92}{$d-1$}\over \scalebox{0.92}{$d-p-1$} }\bigr)\, \oiint  ds^{j} 
 \,    \langle\hskip -1mm \langle   {\cal J}_{j}
\rangle\hskip -1mm\rangle 
\ ; \,}  
 \eea


\smallskip

 ({\footnotesize II}) The norm, $C_D$, of the displacement operator  
\bea\label{def2}
\boxed{\, T_p^{\,(\footnotesize\rm II)} \,  := \, C_D \,   { \pi^{p/2}\, \Gamma( {p\over 2}\scalebox{0.95}{$+1$})\over(p+2)\,  \Gamma(p+1) }\ . \, 
}
\eea

\smallskip
\noindent Here  $\langle\hskip -1mm\langle \cdots \rangle\hskip -1mm\rangle$ denote
 DCFT correlation functions in $\mathbb{R}^d$, 
  the defect spans a   $\mathbb{R}^p$ subspace, 
and the integral is over the (hyper)sphere around the defect in the transverse $\mathbb{R}^{(d-p)}$. 
 The  dilatation current  is ${\cal J}_j = x^m T_{mj}$,  and the
 CFT stress tensor   obeys  the conservation equation 
 \bea\label{conser}
 \partial_m T^{mj}  = \delta^{(d-p)}(x ) D^j
 \eea  
 with  $j$ labelling transverse directions. Eq.\eqref{conser}   fixes unambiguously 
 the norm  of the displacement operator  $D^j$, which  
 becomes a piece of  DCFT data \cite{Billo:2016cpy}. The two definitions are illustrated in figures 2 and 3.   \smallskip  
 
 \setcounter{figure}{1}
 
\begin{figure}[!htbp] 
\begin{minipage}[c]{0.6\linewidth}
\includegraphics[width=\linewidth]{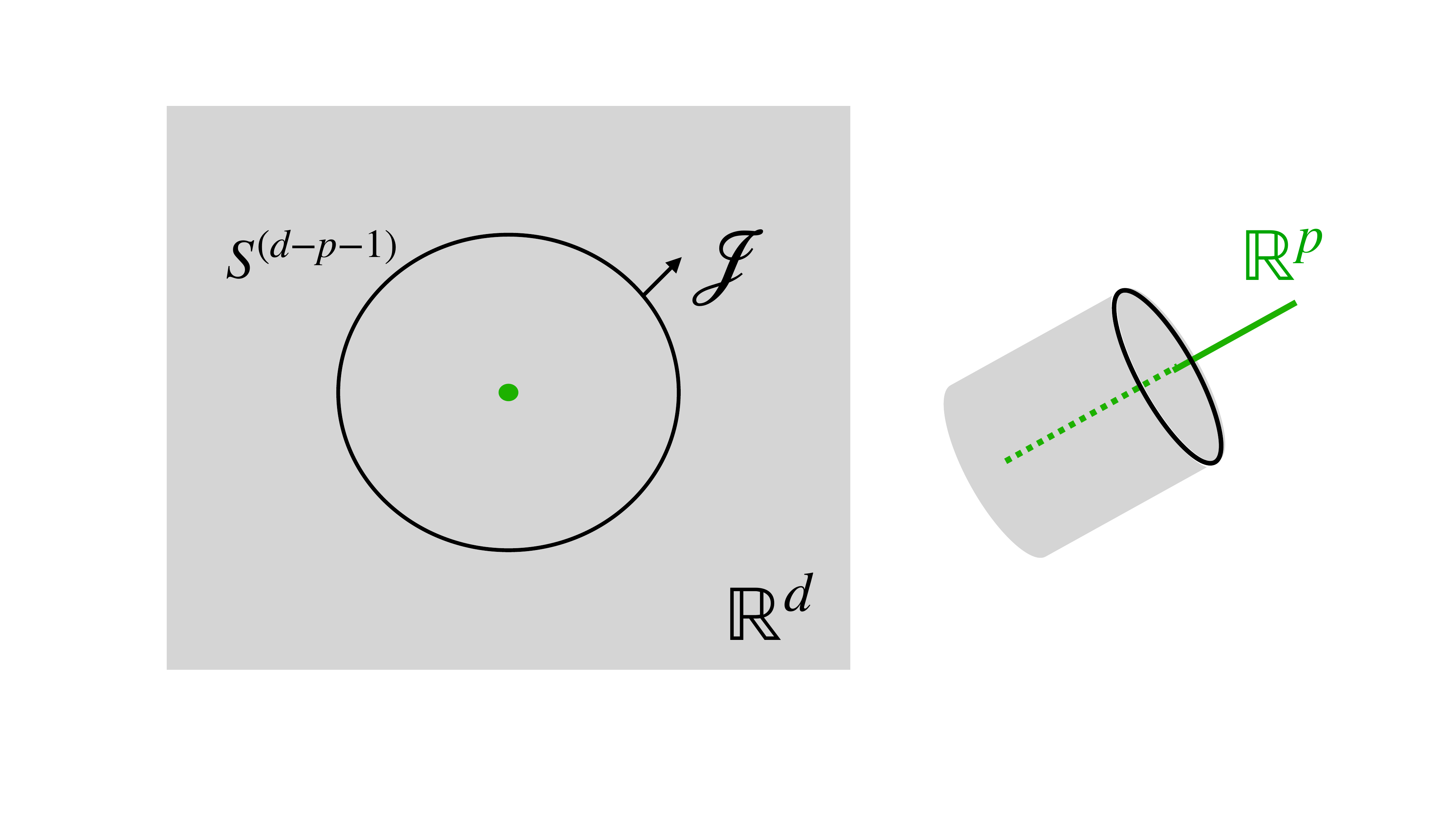}
\vskip -2mm
 \caption{\footnotesize The gravitational tension is\\  the integral of the dilatation current $\cal J$ \\
 over a sphere around  the defect.}
\end{minipage}
\hfill
\begin{minipage}[c]{0.6\linewidth}
\includegraphics[width=\linewidth]{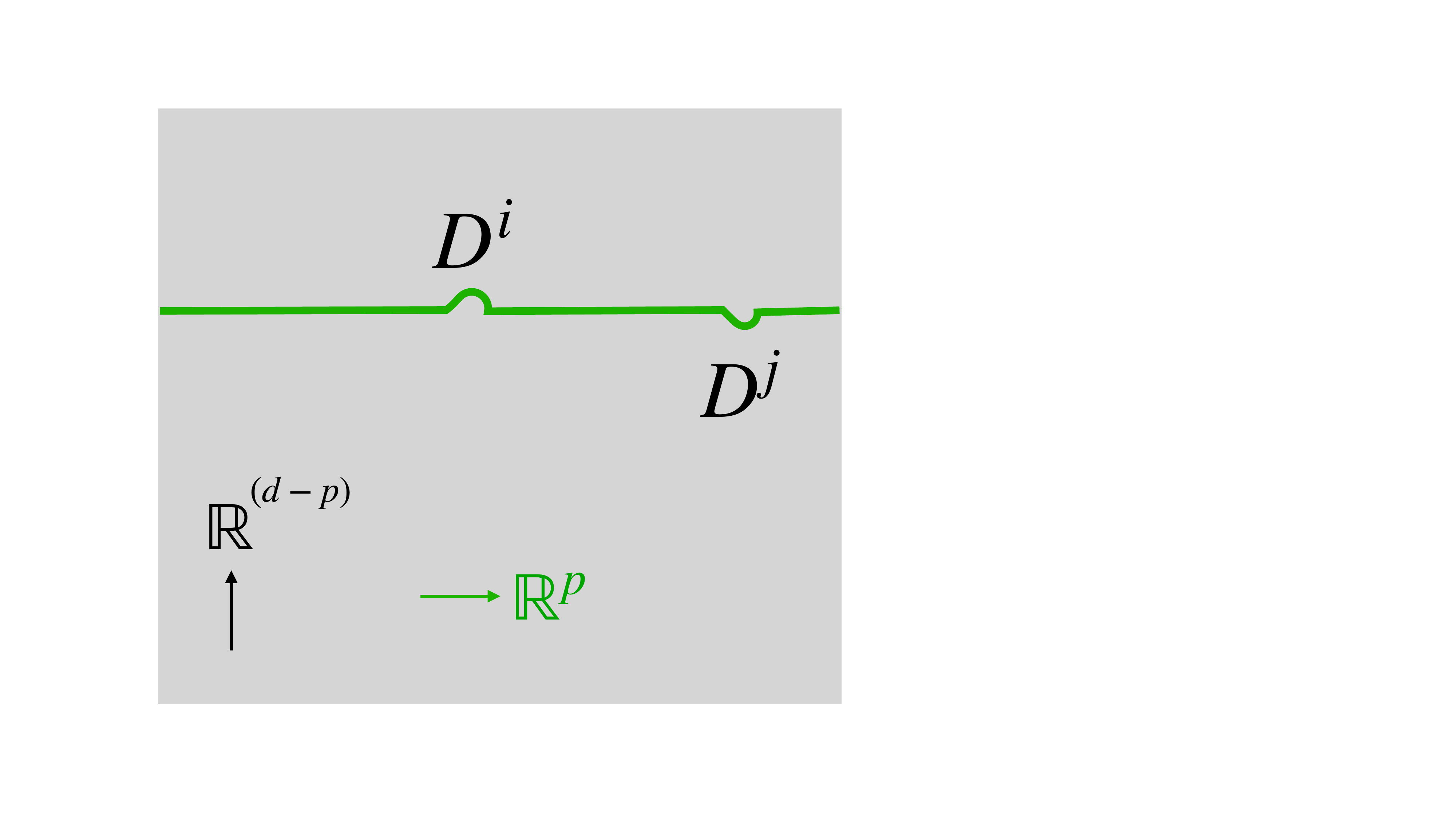}
\vskip -2mm
\caption{\footnotesize The inertial tension, or\\   `stiffness',  is given by the 2-point \\
  \, \, function of the displacement. }
\end{minipage}%
\end{figure}

Definition  ({\footnotesize I}) 
 is the natural extension of  the dilatation charge   $\Delta \simeq m_0$, to which it reduces  for $p=0$. 
 It  is  obtained from  the  graviton one-point function in AdS, so 
  we refer to  it as the  {\it gravitational tension}. 
  Note that in the case of  domain walls both $ \langle\hskip -1mm \langle   {\cal J}_{j}
\rangle\hskip -1mm\rangle$ and $d-p-1$ are zero,  so  the gravitational tension is not defined.

 Definition ({\footnotesize II}), on the other hand, 
   only makes sense  for extended defects ($p>0$). 
It  is a measure of   stiffness, so  we refer to it  as
  {\it inertial  tension}. 
For  line defects ($p=1$)
 it is a multiple of  the   Bremsstrahlung function  \cite{Mikhailov:2003er,Correa:2012at,Fiol:2012sg},
 and in the special case $d=2$, where a line defect is also an interface, it  is proportional 
 to  the  energy reflection coefficient
 \cite{Meineri:2019ycm}. 
Our interest in the  definition of tension
 was  spurred  by holographic calculations in this latter context
\cite{Bachas:2020yxv,Bachas:2021fqo,Baig:2022cnb,Bachas:2022etu},  and 
 part of our motivation for  the present work was to generalise them  to higher dimensions. We will not discuss boundaries  in this work -- viewed as limits
 of interfaces they have infinite tension \cite{Bachas:2020yxv}.
 
     We should  note that  definition  ({\footnotesize I}) is reminiscent of earlier proposals for an invariant tension \cite{Deser:1988fc, Myers:1999psa,Townsend:2001rg,
  Harmark:2004ch,Traschen:2001pb,Kastor:2004jk} which also involve the asymptotic behaviour of the metric. 
  These authors considered  
  spacetimes whose asymptotics were `transverse flat' or  `planar empty AdS',  and they assumed the existence of an (exact or asymptotic)  spacelike 
  Killling isometry. Such an isometry is not required  for our definitions. Although the AdS boundary in  figures 2 and 3  is flat and the defect  linear,  all one really
   needs is that the radius of the sphere   and the separation of the two displacement insertions  be much smaller than all  other DCFT scales.

Our  main technical result  will be  to show, using Witten diagrams, that the right-hand sides of 
 \eqref{def1} and \eqref{def2} reduce to the bare   Nambu-Goto tension
 for classical  probe branes. In this calculation  normalisation factors matter  and are subtle. 
 We   will fix them  by verifying   that the two-point function 
 $\langle\hskip -1mm \langle T^{mn} \,{\cal D}^{k} \rangle\hskip -1mm\rangle$
 obeys, in the above limit,  the 
  Weyl and diffeomorphism Ward identities \cite{Billo:2016cpy} of the (putative) dual DCFT. 

 For   
  defects of dimension 
 $p=2$ or  $4$ the displacement norm and  one-point function of the stress tensor are related 
 to  Graham-Witten anomalies, see \cite{Chalabi:2021jud} for an updated  summary.
  In the Nambu-Goto limit these  can be obtained from the renormalised volume (alias `Willmore energy') of minimal submanifolds 
 \cite{Graham:1999pm,Graham:2017bew}. 
 The use of Witten diagrams is a   lowbrow method for computing  such volumes for all  $p$ and $d$.\footnote{\,It also allows in principle the systematic calculation of
 quantum corrections.}


 The inertial and gravitational tensions are not,  in general,  equal 
 beyond the thin-classical-probe limit. 
  But
  for certain  Wilson-line  and surface defects in $d=4$   theories,  superconformal Ward identities 
     \cite{Bianchi:2018zpb,Bianchi:2019sxz} can be used to show 
  that  
  $T_p^{\footnotesize\rm (I)}=T_p^{\footnotesize\rm (II)}$ exactly. 
    We  conjecture  that this should be true for   large classes of  superconformal defects.  
A similar  
conjecture was put forth in ref.\cite{Bianchi:2019sxz} by extrapolating  some known  results  to arbitrary  $p$ and $d$.
We will see that the  two conjectures are actually  identical.




  The rest of the paper is organised as follows.   In section \ref{sec:2} 
   we   
   show using Witten diagrams  that \eqref{def1} and \eqref{def2}
reduce to the  Nambu-Goto tension for thin classical probes.  We  assume a  natural but adhoc normalisation for the  displacement source.
   In   section \ref{sec:3} we  calculate in the same limit  the  two-point function 
 $\langle\hskip -1mm \langle T^{mn} \,{\cal D}^{k} \rangle\hskip -1mm\rangle$,  and show that it obeys the DCFT Ward identities    \cite{Billo:2016cpy} which  relate  it to 
the  displacement norm $C_D$  and to
$\langle\hskip -1mm \langle T^{mn} \rangle\hskip -1mm\rangle$. 
This confirms the normalisations of  the previous section. 
  In  section \ref{sec:4} we consider  some specific defects,  check that our calculations are
    consistent with   results  obtained  by other  methods,  and  
 comment on the quantum and gravitational corrections. 
Section \ref{sec:5} explains how supersymmetry can protect 
  the relation $T_p^{\footnotesize\rm (I)}=T_p^{\footnotesize\rm (II)}$ from such  corrections\,
 and compares our conjecture to that of ref.\cite{Bianchi:2019sxz}.  This part  deserves further study. 
The constraints on  DCFT correlation functions are  reviewed, for the reader's convenience, 
 in   appendix  \ref{app:A}. 




 \section{Nambu-Goto probes}\label{sec:2} 
 
 To  show that the right-hand sides of  \eqref{def1} and \eqref{def2} 
  reduce to the  Nambu-Goto tension  for thin classical probes  
we will use Witten diagrams  \cite{Witten:1998qj,Freedman:1998tz}.  These have been applied to  DCFT \cite{Giombi:2017cqn,Rastelli:2017ecj,Karch:2017wgy,Giombi:2018hsx,Bianchi:2020hsz,Bliard:2022xsm,Goncalves:2018fwx,Gimenez-Grau:2023fcy} 
mostly in order  to compare to  exact results
(from localisation and integrability) for
 superconformal Wilson lines. 
Our calculations are simpler but the  focus   is different.  We are  mainly concerned with the  normalisation of bulk and defect sources.

 
 

\subsection{Inertial tension}\label{sec:2.1} 
  
  We begin with the definition \eqref{def2}. 
  The norm   of the displacement operator   is  read from the two-point function\footnote{Double brackets will
  always stand for normalised correlation functions   in the background of the defect,  both for bulk and  for defect operators.
  }
   \vskip -6mm
\bea
\langle\hskip -1mm \langle   D^i (\vec\tau) D^j (\vec\tau^{\,\prime}) \rangle\hskip -1mm\rangle\ =\ 
{C_D\,\delta^{ij}  \over \vert \vec \tau-\vec \tau^{\,\prime}\vert^{\,2p+2}}\   , 
\eea
where $\vec \tau$ and $\vec \tau^{\,\prime}$ are points on the defect. Both the norm of $D^j$ and
its  scaling dimension, $\Delta = p+1$,  are part of the DCFT data. 
 The first task is to compute them on  the gravity side.

  

 The metric of  Euclidean AdS$_{d+1}$   in Poincar\'e coordinates is
  \bea
  ds^2 = {\delta_{\mu\nu}\,dy^\mu dy^\nu \over (y^0)^{2} } \qquad {\rm with} \quad \mu, \nu =0, 1, \cdots , d\ \ . 
  \eea  
   The static  brane sits at $y^j=0$ for $j=p+1, \cdots ,d$\,. 
  It breaks the AdS$_{d+1}$  isometries    to SO\scalebox{0.9}{$(1, p+1)\times$}{\rm SO}\scalebox{0.9}{$(d-p)$},   but
 this  residual symmetry does not act simply on the  coordinates $y^\mu$. 
 As realized in
    \cite{Giombi:2017cqn},  the  system of coordinates in which the residual isometries are  manifest  
    is\,\footnote{ Here and in what follows we
use letters from the beginning of the Greek and Latin alphabets for the directions along the brane, respectively the defect; 
  early middle Latin  letters ($i, j, k$)
  for the transverse directions;   and late middle Greek and Latin letters for all AdS directions, respectively those of the AdS  boundary.
 Context will hopefully help avoid confusion. A quick mnemonic is the Greek/Latin correspondence
  $x^\mu = (x^0, x^m)$ and $x^\alpha = (x^0, x^a)$.  
       }
$$
 y^0 \, = \, x^0 \, \Bigl( {1-  {1\over 4} \vert x_\perp\vert^2  \over 1+ {1\over 4}\vert x_\perp\vert^2 } \Bigr)\ ,\quad   y^a =  x^a\quad {\rm for }\quad  a= 1, \cdots , p
$$\vskip -5mm
\bea\label{goodcoordinates}
\quad \hskip -3mm  \quad {\rm and} \quad y^j \, = \, x^0 \, \Bigl( {x^j  \over 1+ {1\over 4}\vert x_\perp\vert^2 } \Bigr)  \quad {\rm for }\quad j= p+1,\cdots , d\   
\eea
with $\vert x_\perp\vert^2 = \sum_j x^j x^j$. The metric in these coordinates  reads
\bea
 ds^2 \,=\, { \delta_{\alpha\beta}\, dx^\alpha dx^\beta \over (x^0)^2}
 \Bigl( {1+  {1\over 4} \vert x_\perp\vert^2  \over 1- {1\over 4}\vert x_\perp\vert^2 } \Bigr)^2 \, +\, { \delta_{ij}\, dx^i dx^j \over (1- {1\over 4}\vert x_\perp\vert^2)^2}\  .
\eea 
 The  unbroken    SO\scalebox{0.9}{$(1, p+1)$}  acts now only on the   $x^\alpha$, not  the  $x^j$. 

 \smallskip
    
 Let     $X^\mu(\tau )$  be the  embedding of the $p$-brane in   AdS$_{d+1}$. The
 Nambu-Goto action is proportional to the bare tension $T_p^{(0)}$, 
 \bea\label{NG}
 I_{\rm NG}\, =\, T_p^{(0)}\hskip -0.5mm \int d^{\,p+1} \tau\, \sqrt{ {\rm det} (\hat g_{\alpha \beta})} \  
 \quad {\rm where} \quad 
 \hat g_{\alpha \beta} = g_{\mu\nu}(X) {\partial X^\mu\over \partial \tau^\alpha }  {\partial X^\nu\over \partial \tau^\beta }   
 \eea
 is the induced world-volume metric. Working in the static gauge,   $ X^\alpha = \tau^\alpha$,    
 and 
 expanding in small transverse  fluctuations
 gives
  \bea\label{NGexpandX}
  I_{\rm NG}= T_p^{(0)} \hskip -0.9mm \int \hskip -0.6mm
  d^{\,p+1} \tau\, \sqrt{ \bar g } \, \Bigl[  1 +  \scalebox{0.95}{${1\over 2}$} \sum_j ( \bar g^{\alpha \beta} \partial_\alpha X^j \partial_\beta X^j  
  + \scalebox{0.90}{$(p+1)$} X^j X^j )
  + \cdots 
 \Bigr] \, .  
\eea
Here $\bar g_{\alpha \beta} = \delta_{\alpha \beta}/(\tau^0)^2$ is the metric of AdS$_{p+1}$ and the dots are terms quartic or higher in $X^j$.  
These  contribute quantum  corrections  to the invariant tension  that  are suppressed by inverse powers of $T_p^{(0)}$.
    The action \eqref{NGexpandX} describes $(d-p)$ scalar fields  
    with  mass $m^2 = p+1$ living in AdS$_{p+1}$. 
 From the   relation $m^2 = \Delta(\Delta -p)$ one sees that  the dual operators have $\scalebox{0.92}{$\Delta = p+1$}$, which is  the  expected   dimension of   
   $D^j$  \cite{Giombi:2017cqn}. 
   
   We extract  $C_D$   following the standard  AdS/CFT recipy \cite{Freedman:1998tz}. The on-shell $p$-brane coordinates  are given
   at leading order   by
   \bea\label{onshellX}
  X^j(\tau^0, \vec\tau) = \int d^p \tau^{\prime} \, K_\Delta (\tau^0, \vec \tau\,;\,  {\vec \tau^{\,\prime} })\,X^j(0, \vec \tau^{\,\prime})\ , 
\eea
  where $\vec\tau,\vec \tau^{\,\prime} $ are $p$-component vectors
 and 
      \bea
K_\Delta (\tau^0, \vec \tau\,;\,  {\vec \tau^{\,\prime} })\,  =\,  
{\Gamma(\Delta) \over \pi^{p\over 2} \Gamma(\Delta- {1\over 2}p)}\, \biggl[ {\tau^0\over (\tau^0)^2+ \vert\vec \tau - \vec \tau^{\,\prime}\vert^2}
\biggr]^\Delta\  
\eea
is the usual  bulk-to-boundary propagator normalised in order 
to approach 
   $\epsilon^{p- \Delta} \,\delta^{(p)}({ \vec \tau - \vec \tau^{\,\prime}})$  in the limit $\tau^0= \epsilon  \to 0$\,. 
The generating functional of  the DCFT  is found by  inserting \eqref{onshellX} in the Nambu-Goto action, 
  \bea\label{genfun}
   {\cal Z}  = {1\over 2}T_p^{(0)}  {
  \scalebox{0.9}{$ (2\Delta -p)$} 
   \Gamma(\Delta) \over  \pi^{p/2} \Gamma(\Delta- {1\over 2}p)}
    \int \hskip -0.6mm d^p\tau \hskip -0.6mm
    \int  \hskip-0.6mm d^p\tau^{\,\prime}\, X^i  \scalebox{0.9}{$ (0,\vec\tau )$} \, X^j  \scalebox{0.9}{$ (0,\vec\tau^{\,\prime} )$}\, 
    { \delta^{ij}  \over \vert \vec \tau-\vec \tau^{\,\prime}\vert^{\,2p+2}}\,,  
  \eea
with $X^j  \scalebox{0.9}{$ (0,\vec\tau )$}$ the renormalised sources for the dual operators which we take to be the $D^j$.
Eq.\eqref{genfun} differs from the naive on-shell action by an extra factor $(2\Delta -p)/\Delta$ which
will be  important for us here. It 
was  first noticed in homogeneous CFTs  by insisting that the regulated action be consistent with
 conformal  Ward identities  \cite{Freedman:1998tz}, and  it was later shown to arise from the  boundary   limit of  the bulk-to-bulk propagator  
   \cite{Klebanov:1999tb,Giddings:1999qu}. 
     
     Setting  $\scalebox{0.9}{$\Delta = p+1$}$ in \eqref{genfun}
      leads to  the following relation between $C_D$ and the Nambu-Goto  tension
   \bea\label{210CD}
   \boxed{C_D = T_p^{(0)} \, {   (p+2)  \Gamma(p+1)   \over  \pi^{p/2} \Gamma({p\over 2}+1)}\,+ \cdots \   }
  \eea
with  dots standing for the quantum and gravitational corrections that we neglected.\footnote{\,Note   that the leading term  only depends 
 on  the defect dimension $p$,  not on  $d$.} 
  \smallskip 
 Comparing eq.\eqref{210CD}  to  eq.\eqref{def2}
  we see  that  $T_p^{\,(\footnotesize\rm II)} = T_p^{(0)} +\cdots $,  as  announced in the introduction.

  But there are reasons to feel uneasy about this calculation. How do we actually know  that $X^j(0,\vec \tau)$
   is the properly-normalised source of the  displacement operator ? This should be   fixed 
   by the conservation eq.\,\eqref{conser}  
   which relates $D^j$ to $T^{mn}$,  and hence the brane coordinates
  to  the   graviton. The relative normalisation of these latter is  fixed by their  transformation under  diffeomorphisms, 
     $\delta g_{\mu\nu} = \nabla_\mu\xi_\nu + \nabla_\nu\xi_\mu $ and $\delta X^\mu = \xi^\mu(X)$. But  the Fefferman-Graham expansion of the metric 
    is singular in the coordinates $x^\mu$, and the existence of an invariant  regulator 
is not clear. Alternatively, one may argue  following ref.\cite{Freedman:1998tz},  that  the normalisation of the sources in \eqref{genfun} 
is determined by conformal Ward identities in AdS$_{p+1}$. But contrary to  the homogeneous CFTs considered in this reference, the defect  does not have
its proper stress-energy tensor. 

   Despite these doubts eq.\eqref{210CD}   turns out to be correct. 
  We will show this  using  the DCFT  Ward identities   in section \ref{sec:3}.

 

\subsection{Gravitational tension} 


   First  we   turn  to the calculation of the gravitational tension.
This depends on the  one-point function of the bulk  stress tensor  which is  fixed,  up to an undetermined  parameter $a_{\rm T}$,
  by the unbroken conformal symmetry. The result  \cite{Billo:2016cpy} 
(see also appendix \ref{app:A}) reads
  \begin{subequations}\label{26}
\begin{align}
&\langle\hskip -1mm \langle   T^{\,ab} \rangle\hskip -1mm\rangle
\,=\, 
 a_{\rm T}  \, \bigl(  {\scalebox{0.95}{$d-p-1$}
\over d}\bigr)\, 
{\delta^{ab}\over \vert x_\perp\vert^d}
\ , \qquad
\langle\hskip -1mm \langle   T^{\,aj} \rangle\hskip -1mm\rangle\, = \, 0\ ,  \label{26a}
 \\ 
 &\ \quad  \langle\hskip -1mm \langle   T^{\,ij} \rangle\hskip -1mm\rangle\, = \,  a_{\rm T} \,\Biggl[ 
   \,{x^i x^j \over \vert x_\perp\vert^{d+2}}
   \,-\,
 \bigl({  \scalebox{0.95}{$p+1$}\over d }\bigr)
{\delta^{\,ij} \over \vert x_\perp\vert^d} \, 
 \Biggr]\ \ .  \label{26b}
\end{align} 
\end{subequations}
One can verify that  $ \langle\hskip -1mm \langle   T^{\,mn} \rangle\hskip -1mm\rangle$  is traceless  and conserved. 
Positivity of the energy density in the Lorentzian theory requires  $a_{\rm T} $ to be negative.
 From eq.\eqref{26b} 
we find   the radial component of the dilatation current in the transverse space
 \bea
  {x^m  x^j\over \vert x_\perp\vert} \,  \langle\hskip -1mm \langle   T_{\,mj } \rangle\hskip -1mm\rangle\,=\, 
  {a_{\rm T}\over \vert x_\perp\vert^{d-1}} \, \bigl({  \scalebox{0.91}{$d-p-1$}\over d
}\bigr)\ . 
 \eea 
 Integrating over the sphere   in the definition \eqref{def1}   gives 
 \bea\label{TwithaT}
\boxed{ T_p^{\,(\footnotesize\rm I)}\,=\, - a_{\rm T}\,\, 
 { 2 \,\pi^{(d-p)/2}   \scalebox{0.92}{$(d-1)$}  \over 
  \Gamma ( \scalebox{0.92}{${1\over 2} (d-  p)$})\, d}\ . }
 \eea
  This expresses  $T_p^{\,(\footnotesize\rm I)}$ in terms of   the 
  DCFT parameter $a_{\rm T}$.  As already noted in the introduction, $ a_{\rm T} =0$\,
for  defects of codimension-one 
for which  the above definition of tension fails.  
 

    We want now to calculate $a_{\rm T}$
  in the thin-brane limit. To this end    let $g_{\mu\nu} =  \bar g_{\mu\nu}   + h_{\mu\nu}$  with $\bar g_{\mu\nu}$ the metric of
  the bulk AdS. To ensure  a smooth  Fefferman-Graham expansion we
    must use the  Poincar\'e coordinates $y^\mu$ for the graviton. 
The Fefferman-Graham expansion of
defect fields, on the other hand, is  regular   in the coordinates $x^\mu$, whence the difficulties of regularisation mentioned  in the previous subsection.  
  Fortunately at the order at which we will work the change of coordinates \eqref{goodcoordinates} is simple, 
   \bea
  y^\alpha = x^\alpha\,(1+ O(x_\perp^2))  \, \quad {\rm and}\quad 
  y^j = x^0 x^j \,(1+ O(x_\perp^2)) \ . 
  \eea
Thus, up to corrections of order $O(x_\perp^2)$, 
 the  static-gauge parametrisation is unchanged  and we  only need to rescale the transverse brane coordinates, $Y^j = \tau^0\, X^j$.
 Furthermore, the usual ultraviolet cutoffs of AdS$_{d+1}$ and AdS$_{p+1}$ are the same at this leading order, $x^0=y^0=\epsilon$.
Expanding the Nambu-Goto action \eqref{NG} in powers of  $h_{\mu\nu}$  and $Y^j$,  and keeping only the terms of order $O(h)$ and $O(hY^j)$, 
we find 
  \vskip -1mm
 $$
 I_{\rm NG}\ =\  \scalebox{0.95}{${1\over 2}$}\, T_p^{(0)} \int d^{\,p+1} \tau\, \sqrt{  \bar g }\,   \bar g^{\alpha \beta }\,\delta \hat g_{\alpha\beta}  
  $$\vskip -5mm
   \bea\label{NGex}
   {\rm with} \quad\ \ 
 \delta \hat g_{\alpha\beta} =   (1  + Y^j \partial_j ) h_{\alpha\beta} + h_{\alpha j} \partial_\beta Y^j + h_{\beta j} \partial_\alpha Y^j + \cdots\ .
   \eea
 Note that after replacing  $Y^j$ by $\tau^0 X^j$  manifest covariance is lost.
 This is because   the   metric  perturbation transforms non-covariantly   under the  isometries of AdS$_{p+1}$. 
 
\begin{figure}[thb!]
 \centering
 \includegraphics[width=.90\linewidth]{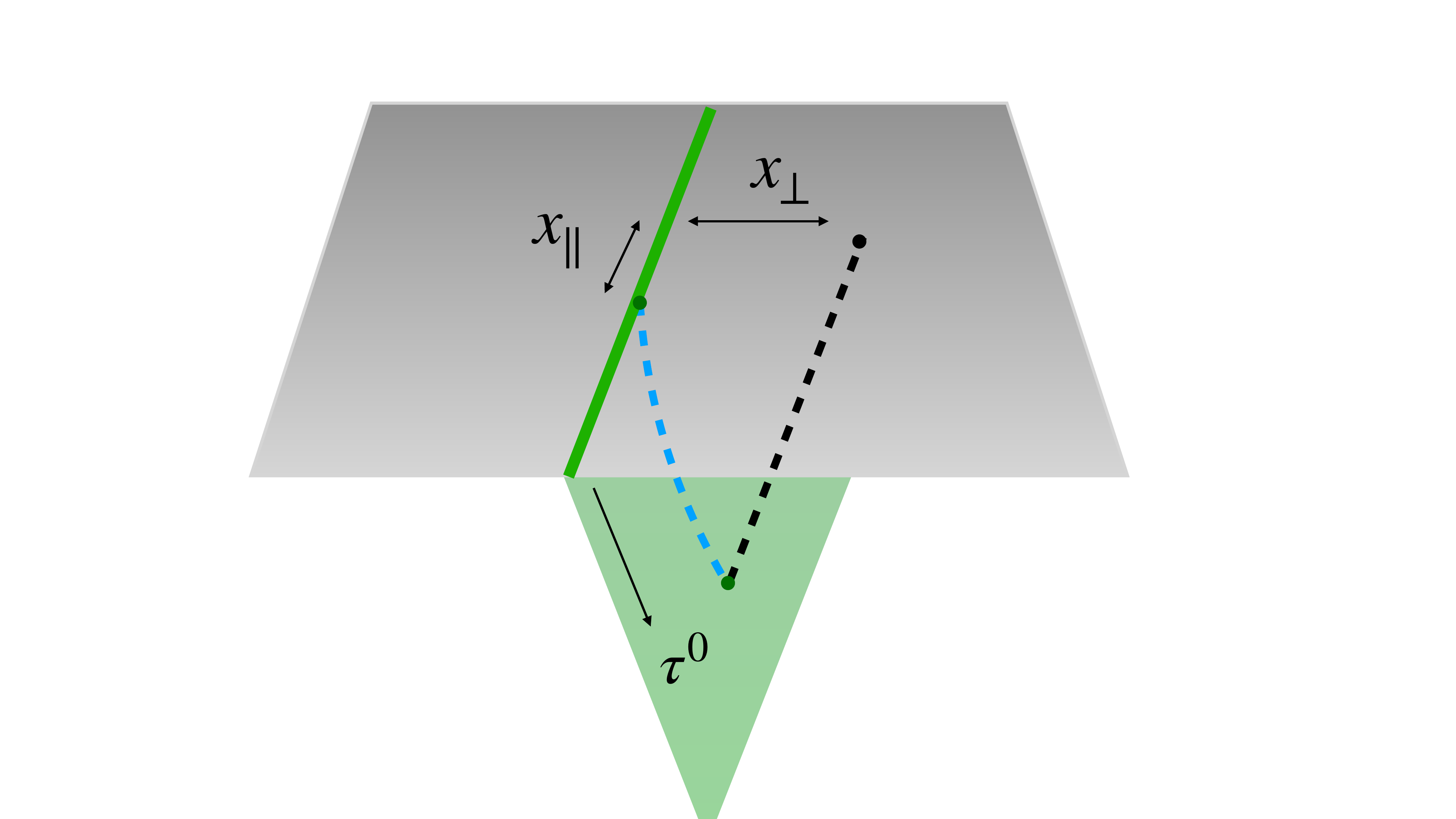}
 \vskip 0.2cm
    \caption{\footnotesize 
    The Witten diagrams calculated in  this and the following  section. The   graviton  
propagator (broken black line) and the displacement-field propagator (broken blue line) stretch from
 the thin brane in the AdS interior to a boundary point in the bulk, respectively on the defect. When the only  external leg is the graviton,  $x_\parallel$ can be set to zero.
   }
 \label{fig:fig2}
  \end{figure}

The leading contribution to $ \langle\hskip -1mm \langle   T^{\,mn} \rangle\hskip -1mm\rangle\,$ comes from the  graviton tadpole 
  in \eqref{NGex}. 
This simplest Witten diagram is shown in figure  \ref{fig:fig2}.  It has a bulk-to-boundary graviton propagator integrated over the world-volume of the static  $p$-brane.
The  graviton  propagator reads \cite{Witten:1998qj,Freedman:1998tz,Polishchuk:2004tm}  
     \bea\label{Kgr}
 K_{\,\mu\nu}^{\, mn}(x,y) \, = \, 
 { C^{[2]}_\Delta \over  (y_0)^{\, 2} }\,  \bigl({y^0  \over    \vert x- y\vert^{2}}\bigr)^\Delta\, 
  J_{\mu r}(x-y) J_{\nu s}(x-y) P^{\,rs; \, mn}\  
  \eea
where  $x= (0, \vec x)$ is  the point on the boundary, 
 \bea\label{CJP}
  C^{[2]}_\Delta = { 2\scalebox{0.95}{$(\Delta +1)$} \Gamma\scalebox{0.95}{$(\Delta-1)$} \over \pi^{d/2}   \Gamma \scalebox{0.95}{$(\Delta-d/2)$}
  }
 \ , \quad
J_{\mu\nu}(z) = \delta_{\mu\nu} - {2 z_\mu z_\nu\over z^2}\   
\eea
 \vskip -4mm
\bea
\ \ \ {\rm and}\quad P^{\,rs; \, mn} \,=\, {1\over 2} (\delta^{rm}\delta^{sn} + \delta^{rn}\delta^{sm}) - {1\over d}\, \delta^{rs}\delta^{mn}   
\ . 
\eea 
Recall that in our conventions  $\mu,\nu\in[0, d]$ are bulk indices whereas  $m,n,r, s\in [1, d]$ are boundary ones.  
 The above propagator is valid for  a spin-2 field of any  mass, the massless    graviton has $\Delta = d$.\,\footnote{\,There
  is a   factor of $2$ missing  in eq.(43) of  ref.\cite{Polishchuk:2004tm}. The
  bulk-to-boundary propagator of a 
   field  dual to an operator of  dimension $\Delta$
 is  $(2\Delta -d)$ 
  times the bulk-to-bulk propagator
  instead of   $(\Delta -d/2)$ as in \cite{Polishchuk:2004tm}. }


  The tadpole in \eqref{NGex},  with the AdS$_{p+1}$ metric $\bar g_{\alpha\beta} =  (\tau_0)^{-2}\, \delta_{\alpha\beta}$, 
gives
\bea\label{hpoint}
 \langle\hskip -1mm \langle   T^{\,mn}(x) \rangle\hskip -1mm\rangle\, =
  \, - {T_p^{(0)} \over 2}  \int {d^{p+1} \tau\over (\tau_0)^{p-1}}\,   \delta^{\alpha\beta} K_{\,\alpha\beta}^{\, mn}(x, \tau)  \, . 
\eea
  Without loss of generality  we take the boundary point to be  at   $x^a=0$. The point in the interior lies on the brane at  $y^\mu =(\tau^\alpha, 0)$. 
 We 
perform  the index contractions in two steps. First
 \bea
 \delta^{\alpha\beta}J_{\alpha r}(x-y) J_{\beta s}(x-y) \, =\,  \delta_{rs}^{\,\parallel}  + {4z_rz_s \, \vert \tau\vert^2 \over \vert z\vert^4} + 
 {2 (\tau_rz_s + z_r \tau_s)   \over \vert z\vert^2} \ , 
 \eea
where $\delta_{rs}^{\,\parallel} $ is the Knonecker symbol in the subspace spanned by the defect,    $z^\mu =x^\mu-y^\mu 
= (-\tau^\alpha, x^j)$ and $\vert z\vert^2 = \vert \tau\vert^2 + \vert x_\perp\vert^2$. 
Contracting next with the tensor $P^{\,rs; \, mn}$ and noting that  $\delta^{rs} z_r z_s  = \vert z\vert^2 - (z_0)^2$ gives\\
 $$
  \delta^{\alpha\beta}J_{\alpha r} J_{\beta s} \,P^{\,rs; \, mn}\,=\, 
 \delta^{\,mn}_{\,\parallel}  + {4z^m z^n  \, \vert \tau\vert^2 \over \vert z\vert^4} + {2 (\tau^m z^n + z^m \tau^n)   \over \vert z \vert^2} 
$$ \vskip -4mm
\bea \label{tensorstr}
\quad\quad\quad  
- {1\over d}\,\delta^{mn} \Bigl[ \,p + {8 \vert \tau\vert^2 \over \vert z\vert^2} - {4 (\tau_0)^2 \vert \tau\vert^2 \over \vert z\vert^4} -  {4 (\tau_0)^2   \over \vert z\vert^2}
\Bigr]\ . 
\eea
For the transverse components $\langle\hskip -1mm \langle   T^{\,ij }  \rangle\hskip -1mm\rangle$ the above tensor structure reads
\medskip
\bea\label{xixj}
{4x^i x^j  \, \vert \tau\vert^2 \over \vert z\vert^4} - {1\over d}\,\delta^{ij} \Bigl[ \,p + {8 \vert \tau\vert^2 \over \vert z\vert^2}
- {4 (\tau_0)^2 \vert \tau\vert^2 \over \vert z\vert^4} -  {4 (\tau_0)^2   \over \vert z\vert^2}
\Bigr]\ . 
\eea
The mixed  components  $\langle\hskip -1mm \langle   T^{\,a j}  \rangle\hskip -1mm\rangle$  vanish by $SO(p)$  symmetry,  and  
in  computing  $\langle\hskip -1mm \langle   T^{\,a b}  \rangle\hskip -1mm\rangle$ one   can  replace for the same reason
$$\tau^a \tau^b\ \to\  {\delta^{ab} \over p} \, \vert \vec \tau\vert^2
%
\ . $$
The integrand 
in eq.\eqref{hpoint}  thus only depends   on $\vert \vec \tau\vert^2$ and $\tau_0$.

 It is simplest to extract the DCFT parameter  $a_{\rm T}$  by looking at  the term  proportional to $x^i x^j$. 
Inserting \eqref{xixj} 
in \eqref{hpoint} and comparing with the   one-point function \eqref{26b} gives
\bea
 - { a_{\rm T}\over \vert x_\perp\vert^{d+2}}
 \,=\, {2\,T_p^{(0)} }\,C^{[2]}_d  \int {d^{p+1} \tau\over (\tau_0)^{p-d+1}}\,   { \vert \tau\vert ^2 \over (\vert \tau\vert^2+ \vert x_\perp\vert^2)^{d+2} } \ . 
\eea
The integral  is computed using Schwinger's trick as follows:
 $$
 -{ a_{\rm T}\over \vert x_\perp\vert^{d+2}}
 \,=\, 
 {2\,T_p^{(0)} }\,C^{[2]}_d 
 \int {d\tau_0\, d^{\,p} \tau \over (\tau_0)^{p-d+1}}\,  {  (\tau_0)^2 + \vert\vec \tau\vert^2 \over \bigl( (\tau_0)^2 + \vert\vec \tau\vert^2+ \vert x_\perp\vert^2\bigr)^{d+2} }  
$$
$$
=  \, {2\, T_p^{(0)} \, C^{[2]}_d   \over \Gamma (\scalebox{0.93}{$d+1$})} 
  \int ds\, s^{d}\bigl( 1 - { \vert x_\perp\vert^2\, s \over  d+1}\,
  \bigr)
  \, e^{-s \vert x_\perp\vert^2} \hskip -1.3mm 
  \int {d\tau_0\, d^p \tau \over (\tau_0)^{p-d+1}}\,  e^{-s (\tau_0)^2 -s \vert\vec \tau\vert^2}   
$$
 $$
=  \, {  T_p^{(0)} \, C^{[2]}_d \over   \Gamma (\scalebox{0.93}{$d+1$})} \, \Gamma \scalebox{1.16}{$({d-p\over 2})$} \pi^{p/2} 
  \int ds \, s^{d/2}\bigl( 1 - { \vert x_\perp\vert^2\,  s \over  d+1}
  \bigr)\,
  \, e^{-s \vert x_\perp\vert^2}  
$$
  $$
 =  \, {1 \over \vert x_\perp \vert^{d+2}}
  {  T_p^{(0)}\,C^{[2]}_d \over   \Gamma (\scalebox{0.93}{$d+1$})} \, \Gamma \scalebox{1.06}{$({d-p\over 2})$}
\pi^{p/2} \, {\Gamma \scalebox{1.16}{$({d \over 2}+ 1)$}\,  d\over 2(d+1)}\ . 
$$
 
 \noindent  Inserting the expression for $C^{[2]}_d$ from \eqref{CJP} leads after a little   algebra  to
  \bea\label{224aT}
 \boxed{ - a_{\rm T}\, =\,  
 { T_p^{(0)}\,  \Gamma (\scalebox{1.1}{${d-p\over 2}$})\, d\over 2\, \scalebox{0.92}{$(d-1)$} \,\pi^{(d-p)/2} } + \cdots \  
 }
 \eea
This expresses the DCFT parameter $a_{\rm T}$ in terms of the Nambu-Goto  tension in the classical  
thin-brane probe limit.  
 Comparing to eq.\eqref{TwithaT} we see that   $T_p^{\,(\footnotesize\rm I)} = T_p^{(0)}+ \cdots$  as announced.
 
 The  terms  in $\langle\hskip -1mm \langle   T^{\,mn} \rangle\hskip -1mm\rangle $   proportional to $\delta^{ab}$ and $\delta^{ij}$
 can be computed similarly and  have  the general form (\ref{26a}, \ref{26b})  dictated 
  by  conformal symmetry. We leave this as an exercise for the reader,  and move on 
 to the   more involved calculation of the  two-point function $\langle\hskip -1mm \langle   T^{\,mn}  \,D^j   \rangle\hskip -1mm\rangle $.




 \section{Ward identities}\label{sec:3}

 We have explained  in section \ref{sec:2.1} why normalising  the displacement operator   in Witten diagrams is tricky. 
We will now  settle the issue  by checking  the DCFT Ward identities that have been derived in ref.\cite{Billo:2016cpy}. 
They are of the form 
 $\langle\hskip -1mm \langle T   D  \rangle\hskip -1mm\rangle\, \sim \, \langle\hskip -1mm \langle T  \rangle\hskip -1mm\rangle\, 
 + \langle\hskip -1mm \langle D D  \rangle\hskip -1mm\rangle\, 
 $\, 
and  fix  therefore unambiguously the normalisations of both $D^j$ and $T^{mn}$ .

  
\subsection{The DCFT   identities}  
  
The   two-point  function 
$\langle\hskip -1mm \langle   T^{\,mn}  \,D^j   \rangle\hskip -1mm\rangle $
is determined by the unbroken conformal and rotation 
  symmetry 
modulo three unknown parameters 
$\mathfrak{b}_{1,2,3}$ (which are called $b_{\rm TD}^{1,2,3} $ in ref.\cite{Billo:2016cpy}). 
 This is  most easily derived in the lightcone formalism as reviewed
in appendix \ref{app:A}. The result is
      \bea\label{physco}  
\langle\hskip -1mm \langle T^{mn}(x)  D^j (0)\rangle\hskip -1mm\rangle
 \,= \,  {
  \vert x_\perp\vert^{p-d}  \over \vert x\vert^{2p+2}} 
\,  F^{mn;\, j}(x)\qquad {\rm with}
\eea
\vskip -3.5mm 
 \begin{subequations}\label{physcorr}
\begin{align}
&F^{ab;\, j}(x) =\,
 {1\over d} [ \scalebox{0.92}{$ (d-p-1)\mathfrak{b}_2 - \mathfrak{b}_1 $} 
  ]\, { \delta^{ab}  x^j } + \,\scalebox{0.92}{$ 4\mathfrak{b}_1$} \, { x^{a}x^{b}x^j \vert x_\perp\vert^2  \over  \vert x\vert^4  }\,
 \,\, ,  \label{physcorr1}
\\
&F^{bi ;\, j}(x) = 
 {
\,   
}   -\mathfrak{b}_3 \, { \delta^{ij}x^b {\vert x_\perp\vert^2 \over \vert x\vert^2}   } 
 +  [ \scalebox{0.92}{$ \mathfrak{b}_3 -2 \mathfrak{b}_1 $} 
  ]\,
 {   x^i x^j x^b \over   \vert x\vert^2 }
+  \scalebox{0.92}{$ 4\mathfrak{b}_1$} \,
 {   x^i x^j x^b\vert x_\perp\vert^2  \over   \vert x\vert^4 }\,
 \,\, ,
\\
&F^{ik;\, j}(x)  =\,
 -{1\over d} [ \scalebox{0.92}{$ (p+1)\mathfrak{b}_2 + \mathfrak{b}_1 $} 
  ]\,
\delta^{ik}  {x^j } +  
 {\mathfrak{b}_3\over 2}    \, \delta^{j(i } x^{k)} \bigl(
  {\scalebox{0.99}{$1$}     } -  
  { \scalebox{0.87}{$2$} \vert x_\perp\vert^2 \over  \vert x\vert^2    }\bigr)
  \nonumber \\
 &\, \hskip 0.2cm  + ( \mathfrak{b}_1 + \mathfrak{b}_2 - \mathfrak{b}_3)
   { x^i x^k x^j  \over    \vert x_\perp\vert^2} 
 +  [ \scalebox{0.92}{$ 2\mathfrak{b}_3 -4 \mathfrak{b}_1 $} 
  ]\,  { x^i x^k x^j   \over     \vert x\vert^2}
 + \scalebox{0.92}{$ 4\mathfrak{b}_1$} \, { x^i x^k x^j \vert x_\perp\vert^2 \over    \vert x\vert^4} \,\,
 .  \label{physcorr3}
  \end{align} 
\end{subequations}
Recall that in our notation $a,b=1, \cdots , p$\, and $i,j,k= p+1, \cdots , d$\,. 
Note also that because of the $D^j$ insertion we  cannot here set    $x^a=0$ as before, see figure \ref{fig:fig2}.
 Thus   $\vert x\vert^2 = \vert x_\parallel\vert^2+ \vert x_\perp\vert^2$. 
\smallskip

    Eqs.\eqref{physco} and \eqref{physcorr} follow from the unbroken   SO\scalebox{0.9}{$(1, p+1)$}$\times$SO$(d-p)$. 
But there are also  constraints coming from the fact that the displacement operator encodes
the action of  the broken   SO\scalebox{0.9}{$(1, d+1)$} symmetries
on the defect. These relate  the one-point function of any primary operator ${\cal O}$ to its two-point function with the displacement
   operator $D^j$, schematically 
    $\langle\hskip -1mm \langle  {\cal O}  D  \rangle\hskip -1mm\rangle\, \sim \, \langle\hskip -1mm \langle {\cal O}  \rangle\hskip -1mm\rangle\, 
 $. 
       When ${\cal O}$ is the stress  tensor  these  identities imply \cite{Billo:2016cpy} 
\bea\label{wardd1}
(p+1) \mathfrak{b}_2 + \mathfrak{b}_1\, = \,    {d\over 2} \mathfrak{b}_3    \ , \quad
\mathfrak{b}_3 = 2^{\,p+2} \pi^{-(p+1)/2}\Gamma ( { \scalebox{0.92}{$p+3$} \over 2} )\,a_{\rm T}\ . 
\eea
The remaining free parameter is 
determined by  the conservation   \eqref{conser} which relates a specific combination of  $\langle\hskip -1mm \langle  T  D  \rangle\hskip -1mm\rangle\,$ to
$\langle\hskip -1mm \langle  D  D  \rangle\hskip -1mm\rangle\,$. This gives \cite{Billo:2016cpy} 
\bea\label{wardd2}
2p\,\mathfrak{b}_2 - \scalebox{0.92}{$ (2d-p-2) $} \, \mathfrak{b}_3\ =  \ { (d-p) \Gamma({d-p\over 2})\over
\pi^{(d-p)/2 }}\, C_D
 \ . 
\eea
Together eqs. \eqref{wardd1} and \eqref{wardd2} can be used 
to express $\mathfrak{b}_1, \mathfrak{b}_2$ and $\mathfrak{b}_3$ in terms of the DCFT data $a_{\rm T}$ and $C_D$.
We want   to verify that these relations are satisfied  when the AdS fields sourcing  $T^{mn}$ and $D^j$ are normalised as in section \ref{sec:2}.


\subsection{The gravity calculation}  
   
       The   leading-order  Witten diagram  for $\langle\hskip -1mm \langle  T  D  \rangle\hskip -1mm\rangle\,$
      has one bulk-to-boundary propagator  for the graviton and
    one for the displacement field. They meet at a quadratic vertex ($\sim hX$) on the $p$-brane, as shown in  fig.\ref{fig:fig2}. Using  the  
    vertices in  \eqref{NGex} and recalling  that $Y^j = \tau^0 X^j$ gives  
  $$
\hskip -1mm  \langle\hskip -1mm \langle   T^{\,mn}(x)D_j(0) \rangle\hskip -1mm\rangle\, =
  \, - {T_p^{(0)} \over 2}  \int {d^{p+1} \tau\over (\tau^0)^{p-1}}\, \delta^{\alpha\beta} 
  \Bigl[   \tau^0\, K_{p+1} (0, \tau )  {\partial\over \partial w^j} K_{\,\alpha\beta}^{\, mn}(x, w)\bigl\vert_{w=\tau}  
 $$\vskip -5mm
\bea\label{hDpoint}
+  K_{\,\alpha j }^{\, mn}(x, \tau) {\partial\over \partial \tau^\beta}(\tau^0 K_{p+1} (0, \tau ) )  \,+\,
K_{\,\beta j }^{\, mn}(x, \tau) {\partial\over \partial \tau^\alpha}(\tau^0 K_{p+1} (0, \tau ) )
 \Bigr]\ . 
\eea
Here $K_{p+1} (0, \tau )$ is the  scalar propagator for   weight $\Delta = p+1$, 
and  in the top line $w^\mu$ should be  
 set equal to $(\tau^\alpha, 0)$  only after having taken the derivative of the propagator in the direction $j$. 

   The computation of the above diagram  is straightforward but tedious.  Below  we   present a sample
   calculation  of  the $x^i x^j  x^k$ terms  in \eqref{physcorr3}. All other terms can be  handled in   the same  way.

The contraction  $\delta^{\alpha\beta}K_{\,\alpha j }^{\, ik}$  was already performed in  eq.\eqref{xixj}. 
Only the first term in this expression will make a contribution 
 proportional to $x^i x^j x^k$. Inserting the  
propagators in \eqref{hDpoint} and replacing $\partial/\partial w^j$ by $-\partial/\partial x^j$ we find 
 \bea\label{start2}
 \braket{T^{\,ik}(x) D^j(0)} \,=\, -4\,T_p^{(0)} \, C^{[0]}_{p+1} C^{[2]}_D \, x^i x^j x^k\,\, {\cal I}(x)\ , 
 \eea 
 where ${\cal I} $ is the   integral
$$
{\cal I}(x)\, =\, \int d^{p+1}\tau\ 
(\tau^0)^{d+1} \, \Biggl[
\frac{(d+2) \, \abs{\tau}^2}{(\abs{\tau}^2+\abs{x_\perp}^2)^{d+3}
\,
((\tau^0)^2 + \abs{\Vec \tau+\Vec x_\parallel}^2)^{p+1}}
 $$\vskip -4mm
 \bea
 \label{integrall}
  + \,
\frac{
(p+1) \, (\abs{\tau}^2 - \abs{\Vec x_\parallel}^2)
- ((\tau^0)^2 + \abs{\Vec \tau + \Vec x_\parallel}^2)
}
{(\abs{\tau}^2 + \abs{x_\perp}^2)^{d+2}
\,
((\tau^0)^2 + \abs{\Vec \tau+\Vec x_\parallel}^2)^{p+2}}\,\Biggr]
\ , 
 \eea 
 and
 \bea
  C_{p+1}^{[0]} =   {\Gamma (p+1) \over \pi^{p/2} \Gamma ({p\over 2}+1)}\,  ,  \quad  
C_{d}^{[2]} =   { 2(d+1) \Gamma (d-1) \over \pi^{d/2} \Gamma ({d\over 2})}
\eea
are the  prefactors in the scalar and spin-2 propagators.  
One can decompose ${\cal I}(x)$
into a  sum of  primitive  integrals
\bea
\, {\cal I}_{A,B,C}
\, = \,
\int \frac{d^{p+1}\tau}{(\tau^0)^{p+1}}\ 
\frac{(\tau^0)^A}{((\tau^0)^2 + \abs{\Vec \tau - \Vec x_\parallel}^2 + \abs{x_\perp}^2)^B\,
((\tau_0)^2+\abs{\Vec \tau}^2)^C}
\eea
$$
=\, \frac{1}{\,\scalebox{0.9}{$\Gamma(B)\,\Gamma(C)$}}
\int  {d^{p+1}\tau \over (\tau^0)^{p+1-A}}
\,{ds \over s^{1-B}}\,{dt\over
 t^{1-C} }
\,e^{-(s+t) (\tau_0^2 + \abs{\Vec \tau}^2)}
\,e^{-s \abs{x_\perp}^2}
\,e^{-st \,\abs{\Vec x_\parallel}^2/(s+t)}\ . 
$$
 \vskip 1mm\noindent  Here we have used  Schwinger's trick, and
 the last exponential in the lower line arises from completing the squares of $\vec\tau$ and $\vec x_\parallel$.  
Doing the $\tau$   integrations  and inserting $1 = \int d\rho\,  \scalebox{0.9}{$\delta(\rho-s-t)$}$ gives 
\vskip -2mm
 $$
 \, {\cal I}_{A,B,C}
\, = \,
\frac{\,\pi^{p/2} \,\Gamma(\frac{A-p}{2})}{2 \,\scalebox{0.9}{$\Gamma(B)\,\Gamma(C)$}}
\int {ds \over s^{1-B}}\,{dt\over
 t^{1-C} }
 \,{d\rho\over 
 \rho^{ A/2 }}
\,e^{-st \abs{\Vec x_\parallel}^2/\rho }
\,e^{- s \abs{x_\perp}^2}
\,  \scalebox{0.9}{$\delta(\rho-s-t)$}\ . 
 $$
 Rescaling the dummy integration variables $s$ and $t$ by $\rho$,  and then doing  the $\rho$ integration gives 
 $$  
 \, {\cal I}_{A,B,C}
\, = \,
\frac{\,\pi^{p/2} \,\Gamma( \scalebox{1.05}{$\frac{A-p}{2}$}) \,
\scalebox{0.9}{$\Gamma(B+C-\frac{A}{2})$}}
{2 \,\scalebox{0.9}{$\Gamma(B)\,\Gamma(C)$}}
\int  
\frac{ds \,dt\ s^{B-1} t^{C-1}}{(st \abs{\Vec x_\parallel}^2 + s \abs{x_\perp}^2)^{B+C-A/2}}\, \delta(1-s-t)
$$ 
  $$  
  \, = \,
\frac{\,\pi^{p/2} \,\Gamma(\scalebox{1.05}{$\frac{A-p}{2}$}) \,
\scalebox{0.9}{$\Gamma(B+C-\frac{A}{2})$}}
{2 \,\scalebox{0.9}{$\Gamma(B)\,\Gamma(C)$}}
\int  dt \, 
\frac{ (1-t)^{A/2-C-1} \, t^{C-1}}{(t \abs{\Vec x_\parallel}^2 +  \abs{x_\perp}^2)^{B+C-A/2}}\,  
$$  
\bea\label{LABC}
\hskip -1mm
 = 
\frac{\,\pi^{p/2} \,\Gamma(\scalebox{1.05}{$\frac{A-p}{2}$}) \,
\scalebox{0.9}{$\Gamma(B+C-$}
\scalebox{1.05}{$\frac{A}{2}$})   \Gamma(
\scalebox{1.05}{$\frac{A}{2}$}\scalebox{0.9}{$-C)$}
}
{2 \,\scalebox{0.9}{$\Gamma(B)\,\Gamma($}
\scalebox{1.05}{${A\over 2}) $} \,  \abs{x}^{2 (B+C)-A}}
\,\, 
{}_2F_1\bigl(
\scalebox{1.05}{$\frac{A}{2}$}-
\scalebox{0.9}{$ C, B+C- $}
\scalebox{1.05}{$\frac{A}{2}, \frac{A}{2},$}\,
 \frac{\abs{\Vec x_\parallel}^2}{\abs{x}^2}
 \bigr)
\,.\,  \,
 \eea
   In the last step  we  used  Euler's representation of the hypergeometric function,   valid 
   under the assumptions $\Re(A-2C) > 0$ and $\Re(C) > 0$.
    For all ${\cal I}_{A,B,C}$  in the decomposition of \eqref{integrall}  these  hypergeometric functions reduce  to simple functions.


Putting together    eqs.\eqref{start2} to \eqref{LABC} gives the $x^i x^j x^k$ term of  $\braket{T^{\,ik}(x) D^j(0)}$. 
After some straightforward algebra the result can be shown to agree with  the lower line of \eqref{physcorr3} for  the following values of the coefficients:
 \beq 
\mathfrak{b}_1 \, = \, \frac{p+2}{2} \,\eta
\ , \qquad
\mathfrak{b}_2 \, = \, \frac{d^2-p-2}{2 (p+1)} \,\eta
\ , \qquad
\mathfrak{b}_3 \, = \, d \,\eta
 \ 
\eeq
 \beq
{\rm with} \quad \,\eta \, = \, - T_p^{(0)} \, \frac{2^{p+1} \,\Gamma(\frac{d-p}{2}) \,\Gamma(\frac{p+3}{2})}{(d-1) \,\pi^{(d+1)/2}}
\ .\qquad
\eeq
It takes a little more algebra to check that the above parameters $\mathfrak{b}_1, \mathfrak{b}_2, \mathfrak{b}_3$ and the  leading-order values of
$C_D$,  $a_{\rm T}$ computed  
in the previous section, eqs.\eqref{210CD} and \eqref{TwithaT},   satisfy the Ward identities   \eqref{wardd1} and \eqref{wardd2}  on the nose. 
This confirms the source normalisations, as announced. 

 As a check of consistency  we  have computed  the full  Witten diagram \eqref{hDpoint}
  and verified that it reproduces the expressions  
    (\ref{physco}, \ref{physcorr}) with the above values of the parameters $\mathfrak{b}_i$.


 \section{Examples}\label{sec:4}  
  
   We consider now some holographic defects whose 
  parameters $C_D$  and $a_{\rm T}$ have been computed by other means. 
 We will check  
  that these reduce to \eqref{210CD} and \eqref{224aT}
  in the appropriate limit,  and  comment on  the quantum ($\sim1/T_p^{(0)}$) and gravitational  ($\sim G_N$) corrections to the invariant tensions.


  
 \subsection{Maldacena-Wilson lines}\label{sec:41} 
  
   The best-studied holographic defect is the half-BPS Maldacena-Wilson line
of ${\cal N}=4$ super Yang-Mills \cite{Maldacena:1998im}.  Its displacement norm is related to many interesting quantities, in particular 
to the Bremstrahlung function  $B(\lambda, N)$  which controls the energy emitted by an accelerating heavy quark. 
Here $N$ is the rank of the gauge group $U(N)$,  and $\lambda$ is the 't Hooft coupling.  
  As  shown in refs.\cite{Correa:2012at,Fiol:2012sg} the following relations hold: 
  \bea\label{CDW1}
  C_D = 12 B =  {6\over \pi^2} \lambda {\partial_\lambda} \log \langle W_{\odot}\rangle \ , 
  \eea
 where $W_{\odot}$ is the circular Wilson loop whose expectation value  was computed exactly as a matrix integral 
  \cite{Erickson:2000af,Drukker:2000rr,Pestun:2007rz} 
\bea\label{CDW2}
\langle W_{\odot}\rangle \,=\,   {1\over N}e^{\lambda/8N}\,  L^{(1)}_{N-1}  \scalebox{1.07}{$(-{\lambda\over 4N})$}\, =\,
  {2\over\sqrt{\lambda} } I_1(\sqrt{\lambda}) + {\lambda\over 48N^2} I_2(\sqrt{\lambda}) + \cdots \ . 
 \eea 
In this expression  $L^{(a)}_n$ are the Laguerre polynomials and $I_1, I_2$ the modified Bessel functions of the first kind. 
 Using supersymmetric localisation \cite{Pestun:2007rz}
one can derive
 similar expressions for  the displacement norm of many other superconformal  line defects in both $d=4$ and $d=3$. 

 In  the planar $(N\to\infty)$ limit one finds
 \bea\label{TFthooft}
C_D \, =\,    {3\sqrt{\lambda}\, I_2(\sqrt{\lambda})\over \pi^2 I_1(\sqrt{\lambda})} \, =\,  {3\sqrt{\lambda}\over \pi^2} 
-{9\over 2\pi^2} + O({1\over\sqrt{\lambda}}) \ . 
\eea
  Now the  relation    between the 't Hooft coupling 
 and the bare  fundamental-string tension  is  $2\pi\,T_F\, \ell^2 = \sqrt{\lambda}\,$, 
 with  $\ell$   the AdS radius that we have set equal to one\,.\footnote{This is derived by expressing  the D3-brane tension and   Newton's constant 
   in terms of
  the string coupling  and $T_F$  \cite{Maldacena:1997re}. Here $T_F = T_1^{(0)}$ is the bare string tension.}  
 Inserting this relation  in  eq.\eqref{TFthooft} 
  gives at the leading order
  $C_D \, =\,      {6 T_F/ \pi} + \cdots\ , 
$
in agreement  with our Nambu-Goto calculation, eq.\eqref{210CD},  for the case  $p=1$. 

   In this example all corrections to the inertial tension defined in eq.\eqref{def2} are known exactly. 
 Quantum fluctuations of the string, in particular, make contributions that
 are down by powers of  $1/T_F \sim  1/\sqrt{\lambda}$,  as expected. 
  They are resummed by the Bessel functions in \eqref{TFthooft}. 
  Gravitational corrections, on the other hand, can be seen from \eqref{CDW2} to be  suppressed  by  powers of $\lambda /N^2$. This is a  milder suppression 
than the naively expected  inverse Schwarzschild radius, 
$r_S^{-1}  \sim (G_NT_F)^{-1}  \sim \sqrt{\lambda}/ N^2$.  
The emergence of this  new scale is due to S duality. It can be understood by noting that  
$${\lambda / N^2} \,\sim \, 1/ \tilde\lambda \, \sim\, {1/T_D}\ , $$
where $\tilde \lambda$ is the 't Hooft coupling of the S-dual gauge theory and $T_D$ is the  D-string tension. When $\lambda, N \gg 1$
the  Compton wavelength of the D-string is larger than the Schwarzschild radius of the F-string, so corrections of  the former to the effective supergravity dominate.


Consider next the gravitational tension defined in terms of  the one-point function of the stress tensor in eq.\eqref{TwithaT}. 
The latter has been  also computed exactly \cite{Okuyama:2006jc,Gomis:2008qa}  with the result    $a_{\rm T}  = -C_D/18$\,.
This linear relation  follows from  superconformal Ward identities \cite{Bianchi:2018zpb}
 which we will discuss in more detail in section \ref{sec:5}. 
 Inserting the above   in eq.\eqref{TwithaT} with $d=4$,  $p=1$ gives
 \bea\label{TF=TF}
 T_F^{\,(\footnotesize\rm I)}\,=\, - 3\pi\, a_{\rm T}\, 
  =  \, {\pi\, C_D\over 6} \,=\,    T_F^{\,(\footnotesize\rm II)} 
 \eea
 where in the last step we used  the definition  \eqref{def2}. Thus the inertial and gravitational tensions of the F-string  in AdS$_5\times$S$^5$ are  
 exactly equal.


\subsection{Interfaces in ${\mathbf d=2}$}\label{sec:42}
 
 The only extended defects in $d=2$ CFTs are line defects. Since these have codimension one,  only inertial tension is
 defined. 
  Integrating eq.\eqref{conser} shows that the displacement operator is in this case   the discontinuity of the stress tensor across the interface. 
 The  norm  of the former can therefore be expressed 
  in terms of the two-point function of the latter  \cite{Billo:2016cpy}. Explicitly\,\footnote{\,The 
extra factor of $4\pi^2$  in this reference comes from a different definition of the  energy-momentum tensor, 
 $T_{ab} = \frac{4\pi}{\sqrt{g}} {\delta S\over \delta g_{ab}}$ as opposed to ${2\over \sqrt{g}}{\delta S\over \delta g_{ab}}$.
 The extra $2\pi$ is convenient in two dimensions, but we use the canonical convention for   arbitrary $d$.
}
 \bea\label{CDsec4}
 C_D =    {1\over2\pi^2} (c_L+c_R - 2c_{LR})\   
\eea
where $c_{L}, c_R$ are the central charges of the CFTs that live on either side of the interface,  and $c_{LR}$ controls the two-point function of the stress tensor
 across the interface \cite{Quella:2006de}. As  shown in 
  \cite{Meineri:2019ycm}, $c_{LR}$ also controls the (universal) fraction of energy transmitted across the interface: $c_{LR}/c_L$ is the transmitted fraction of energy
 for excitations coming from the left, and  $c_{LR}/c_R$ for those incident  from the right.

  Although $C_D$ can be readily computed in many I(nterface)CFT$_2$  models, 
much less is known about it in theories with exact holographic duals.\footnote{\,In the simplest  example of  Janus interfaces,  $c_{LR}$ has been  computed 
in gravity  \cite{Bachas:2022etu} as well as at the symmetric-orbifold point of the conjectured dual CFT \cite{Baig:2023ahz}. But  how to  extrapolate  between these two calculations is unclear.
} 
 One can however compute it in a bottom-up model of a thin  back-reacting 1-brane in AdS$_3$ 
 with the result \cite{Bachas:2020yxv,Bachas:2021fqo}
\bea\label{CLR}
c_{LR}\,  =\, {3\over G_N}  \left( {1\over \ell_L} + {1\over \ell_R} + {8\pi  } G_N T_1^{(0)} \right)^{-1} 
\ ,   
\eea
where $\ell_L$ and $\ell_R$ are the AdS radii on either side of the brane. 
The brane is here treated as classical and thin, but its full back reaction
 is accounted for by solving  Israel matching conditions.

 Taking the zero-tension limit of these conditions
 can be  tricky (for a recent discussion see  \cite{Banerjee:2024sqq}), 
but we   avoid such subtleties by setting $\ell_L=\ell_R = \ell$,  as appropriate for probe branes.  
 Using   the Brown-Henneaux formula
 \cite{Brown:1986nw} $c= 3\ell/2G_N$ and inserting   \eqref{CLR} in \eqref{CDsec4} gives (in units $\ell=1$)
 \bea
 C_D =  {6 T_1^{(0)}/\pi \over 1+ 4\pi G_N T_1^{(0)}} \ \ . 
 \eea
 This agrees with the  Nambu-Goto formula \eqref{210CD} for $G_N\to 0$. Note also  that
 the Israel matching conditions repackage the gravitational screening of the bare tension into a simple denominator.


\subsection{Graham-Witten anomalies }

As one last  check,  
 consider even-$p$  defects which are known to have new Weyl anomalies of both   type-A and  type-B \cite{Graham:1999pm,Schwimmer:2008yh}.
For surface defects ($p=2$)
in particular  there are  three irreducible anomalies
\bea
T_m^m\bigl\vert_{\rm Defect} \, = \, {1\over 24\pi} \bigl(  {\mathbf a}^{(2)} R + {\mathbf d}_1^{\,(2)} \bar K^i_{ab}\bar K_i^{ab} -  {\mathbf d}_2^{\,(2)}W^{ab}_{\ ab} \bigr) \ . 
\eea
Here R is the  Ricci scalar on the defect, $K^i_{ab}$ is the traceless part of
the extrinsic curvature, and $W_{abcd}$ is  the pullback of the bulk Weyl tensor.
 The  coefficients  ${\mathbf d}_1^{\,(2)}$ and ${\mathbf d}_2^{\,(2)}$  are proportional,  respectively, to $C_D$ and to $a_{\rm T}$.\footnote{\,This is consistent 
with the fact that the Weyl tensor vanishes identically in $d=3$
where the surface is an interface and thus $a_{\rm T}$ is   zero.
}   DCFT calculations give
   \cite{Lewkowycz:2014jia,Bianchi:2015liz,Jensen:2018rxu}
\bea
C_D  = {4\, {\mathbf d}_1^{\,(2)} \over 3\pi^2}\ \ \   (d=4)\   ; 
 \qquad
 a_{\rm T}\, =\, - {d\, \Gamma ({d\over 2} \scalebox{0.92}{$ - 1$}) \, {\mathbf d}_2^{\,(2)} \over  12  (d-1)    \pi^{d/2}  }  \ \ \  \  (\forall d>3) \ ,  
\eea
while from the probe-brane holographic calculation of ref.\cite{Graham:1999pm} one finds
 \bea 
   {\mathbf d}_1^{\,(2)} =  {\mathbf d}_2^{\,(2)} = 6\pi  T_2^{(0)}\  
\eea
  for all $d>3$. Eliminating the anomaly coefficients gives  the $p=2$ relations \eqref{210CD} and \eqref{224aT},  which we obtained from Witten diagrams.
Note that
 the mathematical construction of  \cite{Graham:1999pm} fixes the normalisation of the displacement operator   consistently with the Ward identities
 at leading order; 
 but we don't know  if  the cutoff subtleties discussed in sections \ref{sec:2} and  \ref{sec:3} can be incorporated  so as to compute quantum 
 and gravitational corrections. 
Besides giving in one stroke the leading result  for all
values of $p$ and $d$, 
the  expansion in terms of  Witten diagrams   is  presumably  the   first step towards a systematic calculation of  such corrections.

   We can also compare  with the results of  \cite{Chalabi:2021jud} for   $p=4$ defects. 
 There are in this case 22  B-type anomalies,    two of which are proportional  to $C_D$ and $a_{\rm T}$,  as in the  $p=2$ case. 
 We quote   from this reference:
 \bea
 C_D  = -{72\over \pi^4}\,  {\mathbf d}_1^{\,(4)}\ ; \quad
   a_{\rm T}  = {d\, \Gamma({d\over 2}-1)\over (d-1) \pi^{d/2}}\, {\mathbf d}_2^{\,(4)}\ . 
 \eea
 Furthermore  the calculation of the Willmore energy of 5-dimensional  submanifolds  gives in the 
 holographic probe limit \cite{Chalabi:2021jud,Graham:2017bew} 
 \bea
 {\mathbf d}_1^{\,(4)}= -\pi^2 T_4^{(0)}\ ;  \qquad {\mathbf d}_2^{\,(4)}\, =\, - { \pi^2 T_4^{(0)} \over d-4}\ . 
 \eea
By eliminating ${\mathbf d}_1^{\,(4)}$ we recover our  ($d$-independent)  relation \eqref{210CD} for $p=4$. 
Eliminating likewise ${\mathbf d}_2^{\,(4)}$ gives
 \bea
 -  a_{\rm T}  = {d\, \Gamma({d\over 2}-1)\,\pi^2 \, T_4^{(0)} \over (d-1)(d-4) \pi^{d/2}}\,  =
 \,  {d\, \Gamma(\scalebox{1.08}{${d-4\over 2} $})\, T_4^{(0)} \over 2 (d-1)  \pi^{(d-4)/2}}\, 
 \eea 
 which  matches  our eq.\eqref{224aT} for $p=4$ at leading order.


\section{Outlook and a conjecture}\label{sec:5} 
 
 The take-away message of  this paper is that one can define two invariant notions of holographic  tension: gravitational tension,  which is the analog 
 of the ADM mass,  and stiffness or inertial tension.  
 Both reduce to the bare   tension $T_p^{(0)}$  for  classical probes coupled to Einstein gravity, but in general (when they are both defined) 
  they are different.  
  
      These tensions are proportional to  the DCFT parameters $-a_{\rm T}$ and $C_D$,  both of which are
     positive in a unitary theory.\footnote{It was pointed out in ref.\cite{Chalabi:2021jud} that $a_{\rm T}$ is positive for $n$-fold cover boundary conditions. 
     This is consistent with the fact that a surplus-angle defect has negative tension, and suggests that such defects might be pathological. 
     A similar question has been raised in 
    ref.\cite{Betzios:2023obs}.  } 
     They vanish for 
   topological defects, which do not couple to the CFT stress tensor and can be  deformed freely. An interesting question
   that we did not explore  is whether these tensions obey a BPS bound when  
    the defect couples to a conserved $p$-form current, i.e. when the dual brane is charged.

   Another question worth  investigating  is   how  tension behaves under fusion. In the thin-brane model of  section \ref{sec:42} tensions simply
   add up \cite{Baig:2022cnb}, but  we have checked with   $d=2$ free-field interfaces  \cite{Bachas:2007td,Bachas:2012bj,Bachas:2013ora} that the 
   difference between the tension of the fusion
   product and the sum of the constituent tensions can have either sign.  Note however  that in these examples
   the tension   is inertial since
   gravitational tension is not defined for $2d$ line defects. 
   Such issues may  be also relevant for the swampland conjectures that feature  extended objects, see e.g. \cite{Ooguri:2016pdq,Lee:2019wij,Lanza:2020qmt}.  

 Here,  however,  we will conclude by coming back  to the observation of section
 \ref{sec:41}  that the inertial and gravitational tensions of  the  F-string in  AdS$_5\times$S$^5$
are exactly equal at all orders. 

This is related to an  interesting  physics conundrum, as explained by Lewkowycz and Maldacena \cite{Lewkowycz:2013laa}.   
The Bremstrahlung parameter $B=C_D/12$ controls the energy emitted by an  accelerating quark,
while the parameter $a_{\rm T}$ controls the energy collected at a distance.  
 If the  two   were  unrelated, the emitted and collected energies would not be the same. 
Ref.\cite{Lewkowycz:2013laa} attributes this   to the difficulty
of separating  the radiated energy   from the self energy of the quark, and suggests why such a separation might be possible in the case of supersymmetric Wilson lines. 

   Whatever  the  resolution of the conundrum, the technical reason behind the relation \eqref{TF=TF} is understood \cite{Fiol:2015spa,Bianchi:2018zpb}
   and can be sketched as follows.   
    The DCFT Ward identities relate the one-point function of {\it any} primary operator
 ${\cal O}$
   to its two-point function with the displacement operator, schematically $\langle\hskip -1mm \langle {\cal O} \rangle\hskip -1mm\rangle \sim 
   \langle\hskip -1mm \langle {\cal O} D\rangle\hskip -1mm\rangle$. 
   For a scalar operator this determines $ \langle\hskip -1mm \langle {\cal O} D \rangle\hskip -1mm\rangle$
   completely, but for a spin-2 primary  an unknown  parameter remains (see appendix \ref{app:A} for details). 
  When  the spin-2  is the stress tensor, the conservation equation eq.\eqref{conser} gives an extra relation of the form  $\langle\hskip -1mm \langle D D \rangle\hskip -1mm\rangle \sim 
   \langle\hskip -1mm \langle T  D\rangle\hskip -1mm\rangle$ which determines this residual  parameter in terms of $C_D$. 
So in general $ \langle\hskip -1mm \langle T D\rangle\hskip -1mm\rangle$ is completely fixed,  but $C_D$ and $a_{\rm T}$ are unconstrained.
 Suppose however that $T^{mn}$ has a scalar superconformal partner whose two-point function,  as we saw, is fixed by its one-point function.
 Supersymmetry may in this case relate the missing parameter in $\langle\hskip -1mm \langle T  D\rangle\hskip -1mm\rangle$ also to $a_{\rm T}$, 
 and  thus relate this latter  to  the displacement norm  $C_D$. 

   Bianchi and Lemos  \cite{Bianchi:2019sxz} realised that the above argument
   can apply to  superconformal defects other than Wilson lines. 
   They considered half-BPS surface defects ($p=2)$  in $d=4, \, {\cal N}=4$ super Yang-Mills and found that 
    in this case $C_D = -12 a_{\rm T}$.  Using the definitions \eqref{def2} and \eqref{TwithaT}  it is easy to check  that   the dual membranes 
   have $ T_2^{\,(\footnotesize\rm I)}= T_2^{\,(\footnotesize\rm II)} $. Another example are BPS surface defects in $d=6$ for which ref. \cite{Drukker:2020atp}
   found $C_D = -40\pi a_{\rm T}/3$. This   implies  again  $ T_2^{\,(\footnotesize\rm I)}= T_2^{\,(\footnotesize\rm II)} $.
   
    It is then  natural to conjecture that whenever the superconformal Ward identities relate $C_D$ and $a_{\rm T}$, the inertial and gravitational tensions coincide,
\vskip -3mm
\bea\label{conj}
 \boxed{
  T_p^{\,(\footnotesize\rm I)}= T_p^{\,(\footnotesize\rm II)}\,
  }\ \ 
   \ \Longleftrightarrow\ \ \
   C_D\, =\, -a_{\rm T}\,  
  {
2(d-1)(p+2)\Gamma (p+1)
\over  
  d\,\pi^{p-{d/2}}\, \Gamma({p\over 2}+1) \Gamma({d-p\over 2})
  }\ .
 \eea
The authors of   \cite{Bianchi:2019sxz} made a similar guess  by extrapolating  known results on the parameters $C_D$ and $a_{\rm T}$ of superconformal defects.
They proposed that \bea\label{conj1}
  C_D\, =\, h \,\,  
  {
2^{p+1} (d-1)(p+2)\Gamma ({p+1\over 2})\,\pi^{(d-p)/2}
\over  
  (d-p-1)\,\pi^{(p+1)/2}\,   \Gamma({d-p\over 2})
  }\ 
\eea
where $h = -a_{\rm T}\scalebox{0.93}{$ (d-p-1)/d$}$\,.\footnote{\,We thank Lorenzo Bianchi for a communication on this issue.}  
The two conjectures look different, but  they are actually the same.
This follows from the  identity
\bea
{\Gamma(p+1) \sqrt{\pi} \over 2^p \Gamma({p\over 2} + 1) \Gamma({p+1\over 2})} \ =\  1\qquad \forall  \ {\rm integer}\ p\,,
\eea
which one can prove  by using the following identity for integer $k$:
$$
\Gamma(k+ {1\over 2}) = \sqrt{\pi}\, { 1\times 3\times \cdots \times (2k-1)\over  2^k}\,=\, \sqrt{\pi}\, {(2k)!  \over k! \, 2^{2k}}\,=\,  {\sqrt{\pi}\,\Gamma(2k+1)\over 2^{2k}\,\Gamma(k+1)}\ . 
$$
Stating the conjecture as $T_p^{\,(\footnotesize\rm I)}= T_p^{\,(\footnotesize\rm II)}$ is  elegant and economic, but the deeper significance,  if any,  
of this observation is not   clear. 

   As we have seen,  the two tensions  are  equal to the bare tension  in the thin-classical-probe limit, but  both  receive quantum and
gravitational corrections. More generally, they  may depend  on  all  the bulk and defect  DCFT moduli. We expect however the superconformal  Ward identities  
to act homogeneously in the moduli space, as in the examples of refs.\cite{Bianchi:2018zpb,Bianchi:2019sxz}.
So if $T_p^{\,(\footnotesize\rm I)}$ and $T_p^{\,(\footnotesize\rm II)}$ are equal at some point 
they should stay  equal  everywhere.


  
  A first step towards an exhaustive  proof would be  to  classify (possibly along the lines  of  ref.\cite{Agmon:2020pde}) all superconformal DCFTs 
   in which the Ward identities lead to a linear relation betweeen $C_D$ and $ a_{\rm T}$. 
 
      \bigskip

     
 {\bf Aknowledgements}:  We thank Lorenzo Bianchi, 
  Jos\'e Espinosa, Chris Herzog, Andreas Karch, Marco Meineri, Giuseppe Policastro, Miguel Paulos, Irene Valenzuela and  Philine van Vliet for discussions. 
  We are grateful to Philine for her comments on the draft.
     C.B aknowledges the hospitality of  the CERN theory department  during initial stages of this work, and thanks Sergey Solodukhin and 
   all the  participants of the STUDIUM workshop
   in Tours for inspiring  conversations.  
 
 \medskip
 
  {\bf Note added}: It was pointed out to us by Shira Chapman that the considerations of this paper  could be applied to the 
 $p=d-2$  twist  defects,  $\tau_n$, which enter  in the calculation of  R\'enyi entropies \cite{Calabrese:2009qy}. Using the results in
  \cite{Faulkner:2015csl,Bianchi:2016xvf} one finds $T^{\,(\footnotesize\rm I)}\not= T^{\,(\footnotesize\rm II)}$
  for general $n$,  but $T^{\,(\footnotesize\rm I)} = T^{\,(\footnotesize\rm II)}$ in the limit $n\to 1$. 
  This is consistent with the fact that, contrary  to the `cosmic brane'  duals of general twist defects  \cite{Dong:2016fnf}, 
    the Ryu-Takayanagi branes  that calculate  entanglement entropy  \cite{Ryu:2006bv} do have a classical-probe limit.



\appendix

\section{Constraints on DCFT correlators }\label{app:A}

To make the paper self-contained, we derive here the general form of the 
DCFT correlation functions used in the main text. The convenient tool is the embedding-space or light-cone formalism. 
This appendix is based on ref.\cite{Billo:2016cpy} but  we work with tensor indices (as in \cite{Weinberg:2010fx} for the homogeneous case)
rather than   with the
  auxiliary vector as in \cite{Billo:2016cpy}.


In the embedding-space formalism
   $\mathbb{R}^d$ is identified with  the projective $(d+2)$-dimensional lightcone 
\bea
X^MX_M = -X^+X^- + X^m  X_m =0\ , \qquad X^M \sim \lambda X^M\ . 
\eea
Here   $m = 1, \cdots, d$. The relation of $X^M$  to the physical coordinates   is 
\bea\label{Xx}
X^M = X^+ (1, x^2, x^m) \ \ \Longleftrightarrow \  \ 
x^m = X^m/X^+  \ . 
\eea
One may choose the section $X^+=1$,  but  Lorentz transformations of
the embedding space need not respect this gauge. 
\smallskip

  To impose the projection  $X^M \sim \lambda X^M$, one limits attention to tensors in the ambient  space that obey the scaling relation
    \bea\label{scalings}
  T^{M_1 \cdots M_r}(\lambda X) = \lambda^{-\Delta}\, T^{M_1 \cdots M_r}(X)\    
    \eea
for some real $\Delta$. We also  impose the transversality conditions
\bea\label{transverse}
X_{M_1}T^{M_1 \cdots M_r}(X) =  \quad \cdots \quad = X_{M_r}T^{M_1 \cdots M_r}(X) = 0\ . 
\eea
 The physical-space tensor  is  obtained by the pullback
\bea\label{pullback}
\hat T^{\,m_1 \cdots m_r}(x) = (X^+)^{\Delta + r}\,  { \partial x^{m_1}\over \partial X^{M_1}}\cdots { \partial x^{m_r} \over \partial X^{M_r}}\, T^{M_1 \cdots M_r}(X)\ .  
\eea
The pullback is first defined in the ambient space before restricting to the lightcone, 
  i.e. $X^-$ is treated as an independent variable so that
\bea
X^+ { \partial x^{m}\over \partial X^{M}} =   (-x^m, 0 , \delta^m_{\,n})\ . 
\eea
 Since  \eqref{pullback}  is invariant under rescalings $X^M\to \lambda X^M$, the physical-space tensor $\hat T^{m_1 \cdots m_r}$ only depends  on the  coordinate $x^m$ as it should. Furthermore
 $X^M ({ \partial x^{m}/ \partial X^{M}})   =0$, so $\hat T^{m_1 \cdots m_r}$ 
is  unaffected  by  shifts  
\bea\label{shift} 
 T^{M_1 \cdots M_r}( X) \to  T^{M_1 \cdots M_r}( X) + X^{M_1} \Omega^{M_2 \cdots M_r}( X)
\eea
 for any 
tensor $\Omega$ with $r-1$ indices. This explains why  only $d$ out of the $d+2$ components for each index $M$ are physical. 
A straightforward but tedious calculation shows that $\hat T^{m_1 \cdots m_r}$ transforms as a conformal (quasi-)primary tensor field  with scaling dimension $\Delta$ \cite{Weinberg:2010fx}. 


   One can also show that any partial trace  of $\hat T^{m_1 \cdots m_r}$ is proportional to the corresponding   trace of $T^{M_1 \cdots M_r}$. 
Using for instance 
 $\eta_{+-}= -{1\over 2}$ and the transversality condition $X_M T^{M+}=0$ 
 for a  2-index tensor, one finds  
    \bea
   \delta_{mn}\, \hat T^{\,mn} = (X^+)^{\Delta + 2}\, \eta_{MN}\, T^{MN}\ . 
   \eea
Furthemore 
the pullback  \eqref{pullback} preserves
 the (anti)symmetrization  of the ambient-space  tensor, so
 irreducible tensor representations  of SO$(1, d+1)$ descend  to   irreducible representations of SO$(d)$. 
 \smallskip


The formalism is easily adapted to accommodate  
  a  planar  $p$-dimensional defect. The defect  breaks the   symmetry to SO$(1, p+1) \times$ SO$(d-p)$, so 
we  write  $M=(A,i)$
 where 
 $A= +,-, 1, \cdots , p$\,\,   labels the  directions along the defect  lifted to the ambient space  and
   $i = p+1, \cdots , d$ \, labels the transverse directions. 
There are now two   invariant tensors, $\eta^{AB}$ and $\delta^{ij}$. 
Following \cite{Billo:2016cpy} we denote the corresponding inner products $X\circ Y :=  X^i Y^j\delta_{ij} $ and
$X\bullet Y :=  X^A Y^B\eta_{AB} $. Clearly   $X\circ X + X\bullet X =0$ since the ambient space vector $X$ is null.

  Let us  see how the unbroken symmetry constrains  the one-point function of a spin-2 primary 
The most general ansatz allowed by the symmetries and  by the scaling  \eqref{scalings}   is
 \begin{subequations}\label{A9}
\begin{align}
&(X\hskip -1mm \circ \hskip -1mm X)^{\Delta/2}\,
\langle\hskip -1mm \langle T^{AB}\rangle\hskip -1mm\rangle \, = \,  
 a_1\,  \eta^{AB} + a_2\,  {X^A X^B\over X\hskip -1mm \circ \hskip -1mm X} \ , \\ 
&(X\hskip -1mm \circ \hskip -1mm X)^{\Delta/2}\,
\langle\hskip -1mm \langle T^{Ai}\rangle\hskip -1mm\rangle 
  \,  =  \,  a_3\,    {X^A X^i\over X\hskip -1mm \circ \hskip -1mm X} \ ,  \\ 
&(X\hskip -1mm \circ \hskip -1mm X)^{\Delta/2}\,
\langle\hskip -1mm \langle T^{ij}\rangle\hskip -1mm\rangle  
    \, =  \,  a_4\,  \delta^{ij} + a_5\,  {X^i X^j\over X\hskip -1mm \circ \hskip -1mm X} \  , 
\end{align} 
\end{subequations}
where the $a_i$ are arbitrary  coefficients. Transversality,  eq.\eqref{transverse},   implies
\bea
a_1-a_2+a_3=0\quad {\rm and}\quad a_3-a_4-a_5=0\ . 
\eea
Eliminating  $a_2$ and $a_5$ in \eqref{A9} gives
\begin{subequations}
\begin{align}
&(X\hskip -1mm \circ \hskip -1mm X)^{\Delta/2}\, \langle\hskip -1mm \langle T^{AB}\rangle\hskip -1mm\rangle
    \, = \,  a_1\bigl( \eta^{AB} +  {X^A X^B\over X\hskip -1mm \circ \hskip -1mm X}\bigr)
 \, + \, a_3{X^AX^B\over  X\hskip -1mm \circ \hskip -1mm X}\ , 
  \\ 
&(X\hskip -1mm \circ \hskip -1mm X)^{\Delta/2}\, \langle\hskip -1mm \langle T^{Ai}\rangle\hskip -1mm\rangle
    \,  =  \,  a_3\,    {X^A X^i\over X\hskip -1mm \circ \hskip -1mm X} \ , 
   \\ 
&(X\hskip -1mm \circ \hskip -1mm X)^{\Delta/2}\, \langle\hskip -1mm \langle T^{ij}\rangle\hskip -1mm\rangle
     \, =  \,  a_4\,  \bigl( \delta^{ij} -  {X^i X^j\over X\hskip -1mm \circ \hskip -1mm X}\bigr)
 \, + \, a_3{X^iX^j\over  X\hskip -1mm \circ \hskip -1mm X}\ . 
 \end{align} 
\end{subequations}
The terms  proportional to $a_3$ combine to  a contribution 
 $\langle\hskip -1mm \langle T^{MN}\rangle\hskip -1mm\rangle   \propto X^M X^N$ which  can be dropped by using the
shift symmetry  \eqref{shift}. 
Finally the zero-trace condition gives one more relation
\bea
(p+1) a_1 + (d-p-1)a_4 = 0\  \Longrightarrow\ \ a_1 =   \bigl(1- { d  \over p+1}\bigr)\, a_4 \,\, . 
\eea
Thus  the one-point function depends on a single parameter, say 
$a_4$.
\smallskip
  
  
To pull back   to the
 physical stress  tensor,  eq.\,\eqref{pullback}, we use the identities
 \bea\label{idpull}
  {\partial x^a\over \partial X^A}X^A =   {\partial x^a\over \partial X^i}X^i = 0\ , \quad 
 - {\partial x^i \over \partial X^A} X^A =  {\partial x^i\over \partial X^j}X^j  = x^i\  
 \eea
 where  we  separated parallel and transverse indices,  
  $m=(a, i)$, see section \ref{sec:2} for conventions and notation. The result reads
\bea\label{1pt}
\langle\hskip -1mm \langle \hat T^{ab}\rangle\hskip -1mm\rangle   \, =  \, \, { a_1\over \vert x_\perp\vert^{\Delta}}\,  \delta^{ab} 
\, , \qquad 
\langle\hskip -1mm \langle \hat T^{\,ij}\rangle\hskip -1mm\rangle   \, =  \,  \, {a_4 \over  \vert x_\perp\vert^{\Delta}}\, \delta^{ij}  -  {(a_4- a_1)\over \vert x_\perp\vert^{\Delta +2}}\, {x^i x^j  }
 \eea
with $\vert x_\perp\vert^2 = \sum_i  x^i x^i$. As a check one can verify that $\partial_m \langle\hskip -1mm \langle \hat T^{\,mn}\rangle\hskip -1mm\rangle  =0$
if and only if $\Delta =d$\,, the canonical dimension of the stress tensor.  Note that the one-point function vanishes  identically
 in the special case $d=p+1$, i.e. when the defect is an 
  interface or boundary.   Eqs. \eqref{1pt} are the same as \eqref{26}  with the identification 
  $a_{\rm T} =  a_4-a_1=a_4\, d/(p+1)$,  and  with hats dropped from the physical stress tensor.

 
\smallskip

  We turn next to the two-point function  
$\langle\hskip -1mm \langle \hat T^{mn}(x)  D^j (y)\rangle\hskip -1mm\rangle$ and 
 its ambient-space precursor  $\langle\hskip -1mm \langle T^{MN}(X)  D^j (Y)\rangle\hskip -1mm\rangle $\,, 
where $D^j$ is  the displacement operator. 
Since  $Y$ is a point  on the defect,  $Y^j=0$ and hence   $X\hskip -1mm \circ \hskip -0.8mm Y =Y\hskip -1mm \circ \hskip -0.8mm Y=0$. 
The only non-zero scalar products are $X\hskip -1mm \circ \hskip -0.8mm X = (X^+)^2 \vert x_\perp\vert^2$ 
and 
\bea
  X\hskip -1mm\bullet\hskip -0.8mm Y   
= \ -{1\over 2} X^+Y^+ \vert x-y\vert^2\ . 
\eea
   $D^j$ has dimension $p+1$, and we again leave the dimension of the spin-2   free  for now.
  The most general form with the correct scaling symmetries is
\bea
 \langle\hskip -1mm \langle T^{MN}(X)  D^j (Y) \rangle\hskip -1mm\rangle \, = \,  (-2 X\hskip -1mm\bullet\hskip -0.8mm Y )^{-p-1}  (X\hskip -1mm \circ \hskip -1mm X)^{{1\over 2}(p+1-\Delta)}  P^{MNj}
\eea  
 with  $P^{MNj}$ built from  the scale-invariant tensors
 \bea
 \eta^{AB},\quad \delta^{ij}, \quad \mathfrak{X}^M = {X^M\over (X\hskip -1mm \circ \hskip -1mm X)^{1/2}}\ \quad {\rm and}
  \quad  \mathfrak{Y}^A = {Y^A  (X\hskip -1mm \circ \hskip -1mm X)^{1/2} \over (-2 X\hskip -1mm\bullet\hskip -0.8mm Y) }\ . 
 \eea
The most general  eleven-parameter ansatz for this tensor  is
\begin{subequations}
\begin{align}
& P^{ABj} = b_1\, \eta^{AB} \mathfrak{X}^j+ b_2 \, \mathfrak{X}^A\mathfrak{X}^B\mathfrak{X}^j
  + b_3 \, \mathfrak{Y}^{(A}\mathfrak{X}^{B)}\mathfrak{X}^j
  + b_4 \, \mathfrak{Y}^A\mathfrak{Y}^B\mathfrak{X}^j \,, \\
  & P^{iBj} = \, b_5\, \delta^{ij} \mathfrak{X}^B + b_6\, \delta^{ij} \mathfrak{Y}^B + 
  b_7 \, \mathfrak{X}^i\mathfrak{X}^B\mathfrak{X}^j
  + b_8 \, \mathfrak{X}^i\mathfrak{Y}^B\mathfrak{X}^j \,, \\ 
  & P^{ikj} = \, b_9\, \delta^{ik} \mathfrak{X}^j 
  + b_{10} \,  \delta^{j(i} \mathfrak{X}^{k)}  + b_{11} \,   \mathfrak{X}^{i}   \mathfrak{X}^{k}   \mathfrak{X}^{j} 
  \,, 
  \end{align} 
\end{subequations}
where  parentheses denote symmetrization. 
\smallskip

The three shift symmetries 
$\delta P^{MNj} \hskip -1mm \propto \mathfrak{X}^M\mathfrak{X}^N\mathfrak{X}^j$, or
$\mathfrak{X}^{(M}\mathfrak{Y}^{N)}\mathfrak{X}^j$ or
$\mathfrak{X}^{(M} \delta^{N)j}$
allow us to  set   
$b_3= b_7= b_{10}=0$ without affecting   the  physical correlator.  
Transversality, $X_M P^{MBj} = X_M P^{iMj} = 0$,  then requires
\bea
&b_1-b_2+ b_5 = 0\ ,\quad
 -{1\over 2} b_4+b_6+b_8 = 0\ ,  \nonumber\\
&-b_5    -{1\over 2} b_6  =0\ , \qquad   -{1\over 2} b_8+b_9 +b_{11} = 0\ . 
\eea
 Solving    for $b_2, b_4, b_5$ and $b_{11}$  leaves  four  free parameters, 
\begin{subequations}
\begin{align}
& P^{ABj} = b_1\, \eta^{AB} \mathfrak{X}^j + (b_1 -{b_6\over 2} ) \, \mathfrak{X}^A\mathfrak{X}^B\mathfrak{X}^j
+2 (b_6+b_8) \, \mathfrak{Y}^A\mathfrak{Y}^B\mathfrak{X}^j \,, \\
  & P^{iBj} = \, -{b_6\over 2}  \, \delta^{ij} \mathfrak{X}^B + b_6\, \delta^{ij} \mathfrak{Y}^B  
  + b_8 \, \mathfrak{X}^i\mathfrak{Y}^B\mathfrak{X}^j \,, \\ 
  & P^{ikj} = \, b_9\, \delta^{ik} \mathfrak{X}^j 
  +({b_8\over 2} - b_9)\,   \mathfrak{X}^{i}   \mathfrak{X}^{k}   \mathfrak{X}^{j} 
  \,.  
  \end{align} 
\end{subequations}
The zero-trace condition  
\bea\label{0trace}
b_1(p+1) + {1\over 2} (b_6+b_8) + b_9(d-p-1)\,  =\,   0\  
\eea
 eliminates one more leaving three. 
  

To compute the  pullback \eqref{pullback} one needs, 
in addition to \eqref{idpull},  the following pullback  identity (in the $X^+=Y^+=1$ gauge)
\bea
 {\partial x^m\over \partial X^A}Y^A   = y^m-x^m := -z^m\, 
 \,.  
\eea
 The  physical correlation functions read 
 \begin{subequations}\label{physcor}
\begin{align}
&\langle\hskip -1mm \langle \hat T^{\,ab}(x)  D^j (y) \rangle\hskip -1mm\rangle  =\,
 {
 z^j\, \vert z_\perp\vert^{p-\Delta}  \over \vert z\vert^{2p+2}
}\Bigl[ b_1\, { \delta^{ab}   } +2 (b_6+b_8) \, { z^{a}z^{b} \vert z_\perp\vert^2  \over  \vert z\vert^4  }\,
\Bigr]\,, 
\\
&\langle\hskip -1mm \langle \hat T^{\, i b}(x)  D^j (y) \rangle\hskip -1mm\rangle\, =\,
 {
z^b\, \vert z_\perp\vert^{p+2-\Delta} \over \vert z\vert^{2p+4}
}\Bigl[  -b_6 \, { \delta^{ij}   } 
- b_8 {   z^i z^j  \over   \vert z_\perp\vert^2 }
+  2(b_6+ b_8)
 {   z^i z^j  \over   \vert z\vert^2 }\,
\Bigr]\,, 
\\
&\langle\hskip -1mm \langle \hat T^{ i  k}(x)  D^j (y) \rangle\hskip -1mm\rangle =\,
 {
 \vert z_\perp\vert^{p+2-\Delta} \over \vert z\vert^{2p+2}
}\Biggl[ \,
b_9\, \delta^{ik}  {z^j\over  \vert z_\perp\vert^2 } +  
 {b_6\over 2}    \, \delta^{j(i } z^{k)} \bigl(
  {\scalebox{0.87}{$1$}\over  \vert z_\perp\vert^2    } -  
  { \scalebox{0.87}{$2$} \over  \vert z\vert^2    }\bigr)
  \nonumber \\
 &\, \hskip 0.2cm  + (b_1 -{b_6\over 2} + {b_8\over 2}-b_9)
 { z^i z^k z^j  \over    \vert z_\perp\vert^4} 
 - 2b_8 { z^i z^k z^j  \over    \vert z_\perp\vert^2\vert z\vert^2}
 + 2(b_6+b_8) { z^i z^k z^j  \over    \vert z\vert^4} \,
\Biggr]
  \,,  
  \end{align} 
\end{subequations}
where  the  scalar invariants are
$ \vert z_\perp\vert^2  =  z^i z_i$ and 
 $  \vert z\vert^2 =  z^\mu z_\mu= z^\alpha z_\alpha + \vert z_\perp\vert^2 $.  
We may choose the displacement operator insertion at $y=0$, so that $z^m=x^m$.  
The parameters  defined in ref.\cite{Billo:2016cpy} are
 \bea
\mathfrak{b}_1 = {1\over 2} (b_6+b_8)\ ,\quad \mathfrak{b}_{2} =b_1-b_9\ , \quad 
\mathfrak{b}_{3} = b_6\ , 
\eea
where  $b_8$  can be eliminated with the help of condition \eqref{0trace}. 
One can now check that   \eqref{physcor} reduce to the expressions (\ref{physco}, \ref{physcorr}) of the main text.
Note that the   tensor structure in \eqref{physcor} is independent  of  the dimension $\Delta$ of the spin-2 operator. 

   As explained in ref.\cite{Billo:2016cpy}, broken-conformal Ward identities  constraint  the parameters $\mathfrak{b}_j$.  
 The first set of identities is universal, i.e. valid for all primary bulk operators. It translates the fact that 
the displacement operator  encodes the effect of conformal transformations on correlation functions in the presence of a flat defect. 
For a scalar operator ${\cal O}$  this determines completely $\langle\hskip -1mm \langle   {\cal O} D^j      \rangle\hskip -1mm\rangle$
 in terms of the one-point function $\langle\hskip -1mm \langle   {\cal O}    \rangle\hskip -1mm\rangle$. Explicitly
   \bea
\langle\hskip -1mm \langle   {\cal O}_\Delta    \rangle\hskip -1mm\rangle\, =  {a_{\cal O} \over \vert x_\perp\vert^\Delta}\ , \quad 
   \langle {\cal O}_\Delta (x) D^j(y)  \rangle =  {b_{{\cal O} D} \,  z^j \over \vert z_\perp\vert^{\Delta-p} \vert z\vert^{2p+2}}\ ,
 \eea 
 \vskip -5mm
 \bea
{\rm with}\qquad  \Delta  \, a_{\cal O} \,=\,  {
 {\pi}^{(p+1)/2} \over  2^p \,\Gamma ( {1\over 2}(p+1)) 
 }\,  b_{{\cal O} D}\ . 
 \eea
For a spin-2   these identities fix only two of the three parameters $ \mathfrak{b}_i$:
 \bea\label{ward1}
(p+1) \mathfrak{b}_{2} =  \bigl( {\Delta\over 2}  \mathfrak{b}_{3} - \mathfrak{b}_{1} \bigr)\quad {\rm and}
\quad 
\mathfrak{b}_{3} = 2 ({2\over \sqrt{\pi}})^{p+1} \Gamma({p+3\over 2})\, a_{\rm T}\ . 
\eea
The remaining free parameter can  be fixed by supersymmetry 
if the spin-2 has a scalar-primary superpartner and for suitable superconformal defects. 
When the spin-2 is the CFT stress tensor  the integrated conservation 
 eq.\eqref{conser}    gives an extra  constraint \cite{Billo:2016cpy}
\bea\label{ward2}
2p\,\mathfrak{b}_{2}- \scalebox{0.96}{$(2d-p-2)$}\, \mathfrak{b}_{3} \, = \,  (d-p) { {\pi}^{(p-d)/2}} \,\Gamma( \scalebox{1.09}{${d-p\over 2}$} )\, C_{D} \ . 
\eea
 This constraint is special to the  stress tensor and hence requires $\Delta =d$. 
 As noticed  by the authors of \cite{Bianchi:2018zpb,Bianchi:2019sxz},  it can be used to relate $C_D$ to  $a_{\rm T}$ when supersymmetry 
 fixes all three $\mathfrak{b}_j$ in terms of the one-point function. 

 
\bibliography{Tension}{}

\end{document}